\documentclass[usenatbib]{mnras}

\usepackage{graphicx}	
\usepackage{amsmath}	
\usepackage{amssymb}	
\usepackage{soul}
\usepackage[flushleft]{threeparttable}

\usepackage{Custom_Functions}

\defcitealias{kollmeier2014}{K14} 	
\defcitealias{danforth2016}{D16}	
\defcitealias{gaikwad2016}{Paper-I}	
\defcitealias{shull2015}{S15}		

\usepackage{xspace}
\newcommand{\fullcodename}{{\xspace}VoIgt profile Parameter Estimation Routine\xspace}	
\newcommand{\codename}{{\xspace}{\sc viper}\xspace}	
\newcommand{\logNHI}{\,$\log{\rm N_{HI}}$\xspace}   		
\newcommand{\dlogNHI}{\,d$\log{\rm N_{HI}}$\xspace}   		
\newcommand{\GHI}{$\Gamma_{\rm HI}$\xspace}   				
\newcommand{\GTW}{$\Gamma_{\rm 12}$\xspace}   				

\newcommand{\ckpc}{\,ckpc}   	

\newcommand{\cmpc}{\,cMpc}  	
\newcommand{\kmps}{\,km\,s$^{-1}$} 
\newcommand{\pcmsq}{\,cm$^{-2}$}   		

\usepackage[usenames, dvipsnames]{color}
\definecolor{mycolor}{RGB}{0,128,0}
\definecolor{newaddcolor}{RGB}{255,0,255}

\usepackage[normalem]{ulem}

\title[\GHI at $z<0.5$]{\fullcodename (\codename): H~{\sc i} photoionization rate at $z < 0.5$}

\author[Gaikwad et.al]{Prakash Gaikwad$^{1}$\thanks{E-mail: \href{prakashg@ncra.tifr.res.in}{prakashg@ncra.tifr.res.in}}, 
Raghunathan Srianand$^{2}$,
Tirthankar Roy Choudhury$^{1}$,
\newauthor{and Vikram Khaire$^{1}$}
\\
$^{1}$National Centre for Radio Astrophysics, Tata Institute of Fundamental Research, Pune 411007, India \\
$^{2}$Inter-University Centre for Astronomy and Astrophysics (IUCAA), Post Bag 4, Pune 411007, India}

\date{}

\pubyear{2015}

\begin{document}
\label{firstpage}
\pagerange{\pageref{firstpage}--\pageref{lastpage}}
\maketitle


\begin{abstract}
We have developed a parallel code called ``\fullcodename (\codename)'' for automatically fitting the H~{\sc i} Ly-$\alpha$ forest seen in the spectra of QSOs. 
We obtained the H~{\sc i} column density distribution function (CDDF) and line width ($b$) parameter distribution for $z < 0.45$ using spectra of 82 QSOs obtained using Cosmic Origins Spectrograph and \codename.
Consistency of these with the existing measurements in the literature validate our code.
By comparing this CDDF with those obtained from hydrodynamical simulation, we constrain the H~{\sc i} photoionization rate (\GHI) at $z < 0.45$ in four redshift bins. 
The \codename, together with the Code for Ionization and Temperature Evolution ({\sc cite}) we have developed for {\sc gadget-2}, allows us to explore parameter space and perform $\chi^2$ minimization to obtain \GHI.
We notice that the $b$ parameters from the simulations are smaller than what are derived from the observations. We show the observed $b$ parameter distribution and $b$ vs \logNHI scatter can be reproduced in simulation by introducing sub-grid scale turbulence. However, it has very little influence on the derived \GHI.
The $\Gamma_{\rm HI}(z)$ obtained here, $(3.9 \pm 0.1) \times 10^{-14} \; (1+z)^{4.98 \pm 0.11} \;{\rm s^{-1}}$, is in good agreement with those derived by us using flux based statistics in the previous paper.
These are consistent with the hydrogen ionizing ultra-violet (UV) background being dominated mainly by QSOs without needing any contribution from the non-standard sources of the UV photons.
\end{abstract}
\begin{keywords}
cosmological parameters - cosmology: observations-intergalactic medium-QSOs: absorption lines-ultraviolet: galaxies
\end{keywords}

\section{Introduction}
\label{sec:intrduction}
The Ly-$\alpha$ forest absorption seen in the spectra of the luminous distant QSOs is one of the most sensitive tools to study the physical conditions of the intergalactic medium (IGM). In hierarchical structure formation models, Ly-$\alpha$ forest is thought to arise from the fluctuations in the cosmic density field \citep{bi1992,bi1993,bi1997,croft1997,croft1998}. The strength and width of the Ly-$\alpha$ absorption lines can be used to trace the ionization and thermal state of the neutral hydrogen (H~{\sc i}) in the IGM. Thus observations of the Ly-$\alpha$ forest have regularly been used to constrain cosmological and astrophysical parameters related to IGM physics.

Various statistics are used in the literature to constrain cosmological and astrophysical parameters from the Ly-$\alpha$ forest observations. These statistics are broadly divided into two cases. In the first case, Ly-$\alpha$  transmitted flux is treated as a continuous field quantity. In particular, the mean flux, the flux probability distribution function (PDF) and the flux power spectrum (PS) have been used to constrain cosmological parameters such as $\Omega_{\rm m}$, $\Omega_{\rm b} h^2$, $\sigma_8$ and $n_{\rm s}$ \citep{mcdonald2000,phillips2001,trc2001,tegmark2004,viel2004a,viel2004b,mcdonald2005,seljak2006,viel2006a,viel2006b,viel2009}, thermal history parameters\footnote{The thermal state of the IGM is described by the effective equation of state parameterized by the mean IGM temperature ($T_0$) and the slope ($\gamma$).} \citep{zaldarriaga2001,bolton2008,lidz2010,becker2011,calura2012,garzilli2015a} and H~{\sc i} photoionization rate \citep[\GHI ;][]{rauch1997,meiksin2004,bolton2007,mcquinn2011,becker2013,pontzen2014,gaikwad2016}. Constraining such quantities by comparing flux statistics between observations and simulations are relatively easier and are frequently used in the high-$z$ ($2 \leq z \leq 6$) studies. In the second case, Ly-$\alpha$ forest is decomposed into multiple Voigt profiles. The line width distribution function  calculated from Voigt profile fitting  is sensitive to the thermal history and the energy injected by various astrophysical processes in the form of heat and turbulent motions in the IGM \citep{schaye1999,schaye2000,mcdonald2001,dave2001a}.  Similarly, the column density distribution function (CDDF) calculated from Voigt profile decomposition is sensitive to \GHI \citep{cooke1997,kollmeier2014,shull2015,gurvich2016} and cosmological parameters \citep{storrie1996,penton2000,shull2012b}. While statistics based on parameters obtained using Voigt profile fitting are useful in deriving thermal history and equation of state of the IGM, the Voigt profile decomposition is usually a time consuming process. Therefore, a large parameter space exploration in simulations is usually difficult.

To constrain \GHI using CDDF at low-$z$ (i.e., $z<0.5$), the UV spectrograph on space telescope is needed for the Ly-$\alpha$ forest observations. Thanks to Cosmic Origins Spectrograph onboard Hubble Space Telescope (hereafter {\sc hst-cos}) we have good quality observations of the low-$z$ Ly-$\alpha$ forest. These observations are used to place good constraints on \GHI \citep{kollmeier2014,shull2015,gaikwad2016}. Previous measurement of \GHI using CDDF by \citet[][hereafter \citetalias{kollmeier2014}]{kollmeier2014} and \citet[][hereafter \citetalias{shull2015}]{shull2015} are conflicting. \citetalias{shull2015} found a factor $3$ smaller \GHI than \citetalias{kollmeier2014}\footnote{The errorbars on \GHI are not evaluated systematically in either \citetalias{kollmeier2014} or \citetalias{shull2015}.}. Note that both used the same {\sc hst-cos} data  but different cosmological simulations. Recently \citet{gaikwad2016} (hereafter \citetalias{gaikwad2016}) used same {\sc hst-cos} data but different statistics namely flux PDF and flux PS to constrain \GHI with appropriate errorbars\footnote{\citetalias{shull2015} analysis is based on ENZO which uses adaptive mesh refinement (AMR) technique. \citetalias{kollmeier2014} used smoothed particle hydrodynamic {\sc gadget-3} code. Whereas, \citetalias{gaikwad2016} used smoothed particle hydrodynamic {\sc gadget-2} code coupled with Code for Ionization and Temperature Evolution ({\sc cite}) see \S\ref{sec:simulation} for more detail. This allowed us to explore the parameter space in detail.}. In contrast to \citetalias{kollmeier2014} and \citetalias{shull2015}, \citetalias{gaikwad2016} varied \GHI as a free parameter, minimize the $\chi^2$ for the two statistics and calculated appropriate statistical and systematic uncertainties in \GHI using covariance matrices for a range of IGM thermal histories. The \GHI measurements and its evolution obtained in \citetalias{gaikwad2016} is found to be consistent with those from \citetalias{shull2015}.

In this paper we measure the \GHI and associated errors at low-$z$ using CDDF by varying \GHI as a free parameter. The basic idea behind constraining \GHI is to calculate the $\chi^2$ between the observed  CDDF, $f_{\rm obs}({\rm N_{HI}},z)$, and the CDDF, $\overline{f}_{\rm sim}({\rm N_{HI}},z,\Gamma_{\rm HI})$, calculated by modeling the Ly-$\alpha$ forest in a  cosmological simulation. The \GHI corresponding to minimum value of $\chi^2$ (i.e. $\chi^2_{\rm min}$) gives the best fit \GHI whereas the associated statistical error is obtained by calculating the parameter values corresponding to $\chi^2_{\rm min} \pm 1$ \citep{press1992}. The modeling of the Ly-$\alpha$ forest is relatively simple \citep{cen1994,zhang1995,miralda1996,hernquist1996} and well understood as they probe mildly non-linear densities of the IGM ($\Delta \sim 10$ at $z<0.5$, see Table 2 in \citetalias{gaikwad2016}). However, at low-$z$ Ly-$\alpha$ absorption could originate from the extended circumgalactic medium of star forming galaxies. The free parameter \GHI could be degenerate with thermal history parameters such as mean IGM temperature $T_0$ and slope of equation of state $\gamma$ \citep[as shown at $z \geq 2$ by][]{bolton2005,bolton2007,faucher2008c}. 
In \citetalias{gaikwad2016}, we developed a module named ``Code for Ionization and Temperature Evolution ({\sc cite})'' that allows one to probe the wide range of thermal history and \GHI easily. Using {\sc cite}, we varied the thermal history parameters at high-$z$ ($z=2.1$) and evolved the IGM temperature up to $z=0$. As shown in \citetalias{gaikwad2016}, the thermal history parameters at low-$z$ are found to be very similar $T_0 \sim 5000 \; {\rm K}$ and $\gamma \sim 1.6$ even if the initial parameters $T_0$ and $\gamma$ at  $z=2.1$ are quite different. This implies that the \GHI derived at low-$z$ are not very sensitive to the thermal history of the IGM at $z \sim 2$.

In order to obtain the CDDF, each Ly-$\alpha$ absorption is usually decomposed into multiple Voigt profile components. Each Voigt profile is defined by $3$ free parameters i.e., line center ($\lambda_{\rm c}$), H~{\sc i} column density (${\rm N_{HI}}$) and line width parameter ($b$).  The manual Voigt decomposition of the large number ($\sim 10000$) of simulated  Ly-$\alpha$ forest is laborious and time consuming. 
Furthermore, the criteria used to fit the number of components ($N_{\rm Voigt}$) to a given identified Ly-$\alpha$ absorption is subjective and need not be unique. Although there are several Voigt profile fitting codes available in the literature like {\sc vpfit\footnote{\url{http://www.ast.cam.ac.uk/~rfc/vpfit.html}},} {\sc alis} \citep{cooke2014}, {\sc gvpfit} \citep{bainbridge2016} it will be invaluable to have a tailor made automatic module that will identify Ly-$\alpha$ absorption regions and fit them with multiple component Voigt profiles where the best fit parameters of the individual components and the minimum number of required components are determined through objective criterion.

We have developed a parallel processing module ``\fullcodename '' (\codename) to fit the Ly-$\alpha$ forest with multiple Voigt profiles automatically. In \codename, the blended and saturated features are fitted simultaneously with multi-component Voigt profiles. An objective criteria based on information theory is used to find the number of Voigt profiles needed to describe the Ly-$\alpha$ forest.
The parallel and automated nature of \codename allows us to simultaneously fit large number of simulated spectra and to explore a wide parameter space efficiently.
For consistency, we used the same code for analyzing the observed and simulated spectra.
We calculated CDDF by consistently taking into account the redshift path length and the incompleteness of the observed sample. We show CDDF and line width distribution obtained for observed data using \codename matches very well with those from \citet{danforth2016}. 


The paper is organized as follows. In \S\ref{sec:observation} and \S\ref{sec:simulation} we explain the {\sc hst-cos} data and cosmological simulation used in this work. \S\ref{sec:method} describes our module \codename that automatically fits Voigt profiles to the Ly-$\alpha$ forest. 
In \S\ref{sec:results} we match observed CDDF with model CDDF to constrain \GHI in $4$ different redshift bins defined in \citetalias{gaikwad2016}. Finally we summarize our results in \S\ref{sec:summary}.
Throughout this work we use flat $\Lambda$CDM cosmology with parameters $(\Omega_{\Lambda}, \Omega_{\rm m}, \Omega_{\rm b}, h, n_s,\sigma_8,Y) \equiv (0.69, 0.31, 0.0486, 0.674, 0.96, 0.83, 0.24)$ reported by \citet{planck2015}. All the distances are expressed in comoving co-ordinates unless and otherwise mentioned. The H~{\sc i} photoionization rate \GHI in units of $10^{-12} \; {\rm s^{-1}}$ is denoted by \GTW.

\section{HST-COS QSO ABSORPTION SPECTRA}
\label{sec:observation}
We used publicly available {\sc hst-cos} science data product\footnote{\url{https://archive.stsci.edu/prepds/igm/}} that consists of a sample of  low redshift Ly-$\alpha$ forest spectra towards 82 UV bright QSOs performed by \citet[][hereafter \citetalias{danforth2016}]{danforth2016}. These QSOs are distributed in the redshift range $z = 0.0628$ to $0.852$. The sample covers the Ly-$\alpha$ forest in the redshift range $0 \leq z \leq 0.48$ with a velocity resolution of $\sim 17 \; {\rm km\:s^{-1}}$ (full width at half maximum). The median signal-to-noise ratio (SNR) per pixel varies from $6$ to $17$ for different sightlines.
  
\citetalias{danforth2016} fitted the continuum to each spectrum and identified several thousand absorption line features that consists of Ly-$\alpha$ lines, higher order Ly-series lines, metal lines from the IGM and the interstellar medium (ISM) of our Galaxy. 
As a part of the {\sc hst-cos} science data product, \citetalias{danforth2016} fitted each absorption feature with multiple component Voigt profiles and provided a table that contains  line identifications (type of the specie and rest wavelength of the transition), redshift of the absorption system, column density, doppler-$b$ parameter, equivalent width (along with associated fitting errors)  and significance level of the absorption line detection. In this work we refer to their Ly-$\alpha$ line catalog as ``\citetalias{danforth2016} line catalog''.

As in \citetalias{gaikwad2016}, we divided the sample into 4 different redshift bins (we denote them by roman numerals I, II, III and IV respectively) ${\overline{z} \pm \Delta z} \equiv (0.1125 \pm 0.0375, 0.2 \pm 0.05, 0.3 \pm 0.05, 0.4 \pm 0.05)$. The size and center of the lowest redshift bin is chosen in a way that avoids the contamination by geo-coronal line emission at $z \le 0.075$. There are total $50,31, 16$ and $12$ lines of sight in the redshift bins I, II, III and IV respectively.  Apart from the intervening Ly-$\alpha$ lines all other lines in the Ly-$\alpha$ forest are treated as contamination in our spectra. We replace all other lines except Ly-$\alpha$ lines by the continuum added with a Gaussian random noise \citepalias[see Fig. 2 in][]{gaikwad2016}. We use these clean spectra for further analysis.

\section{Automatic Voigt profile fitting code}
\label{sec:method}
In this section we describe our automated Voigt profile fitting procedure ``\fullcodename'' (\codename). The same code has been used to fit the observed and simulated Ly-$\alpha$ forest spectra to constrain \GTW in \S\ref{sec:results} from CDDF. The algorithm is broadly divided into 3 steps; first we identify the absorption lines and region bracketing these lines, next in these regions we fit as many Voigt components as necessary based on an objective criteria. In the final step we accept the Voigt profile fit for a line based on a significance level of the fit. We now discuss each step in detail below.
\begin{enumerate}
\item \textbf{Line and region identification : }
Following \citet{schneider1993} and \citetalias{danforth2016}, first we estimate the ``crude significance level'' (hereafter CSL) to identify the lines,
\begin{equation}\label{eq:csl}
{\rm CSL} = \frac{W(\lambda)}{\overline{\sigma}(\lambda)}
\end{equation}
where $W(\lambda)$ and $\overline{\sigma}(\lambda)$ are ``equivalent width vector'' and ``line-less error vector'' respectively. The CSL defined in this way has the advantage that the unresolved features are unlikely to be identified as lines \citep[see][for details]{schneider1993}. $W(\lambda)$ and $\overline{\sigma}(\lambda)$ are obtained by convolving the normalized flux and line-less error (i.e., error on flux if absorption lines were absent), respectively, with a representative line profile (see \S2.3 of \citetalias{danforth2016}). The representative line profile is a convolution of Gaussian (of a Doppler parameter of $b=20 \: {\rm km \: s^{-1}}$) with {\sc hst-cos} line spread function (LSF).  We repeated the procedure with different values of the doppler $b=50,100 \: {\rm km \: s^{-1}}$ to incorporate any missing narrower, broader and blended lines. 
The second panel from the top in Fig. \ref{fig:vpfit-steps} shows  the CSL estimated for the spectrum shown in the top panel.
Initially we identified all the lines with maxima satisfying ${\rm CSL} \geq 1.5$ shown by red stars in second panel of Fig. \ref{fig:vpfit-steps}\footnote{The cutoff used in this work is smaller than that used by \citetalias{danforth2016} (${\rm CSL} \geq 3$). This results in more number of identified lines in  our initial line catalog as compared to \citetalias{danforth2016}. However, in the final step most of the extra identified lines at lower significance level are rejected.}.
Next we find a threshold on either side of the maxima to enclose the line in a region. A minimum is accepted as a threshold if ${\rm CSL < 1.5}$. A minimum with ${\rm CSL > 1.5}$ indicates that the lines are blended hence we search for the next minimum until we meet the condition ${\rm CSL < 1.5}$ to accept it as a threshold.
We then merge the overlapping regions (if any) into one bigger region for blended lines (e.g., see the yellow shaded region in second panel of Fig. \ref{fig:vpfit-steps}). This procedure allows us to identify and fit the blended lines simultaneously.
\InputFigCombine{vpfit_steps.pdf}{175}
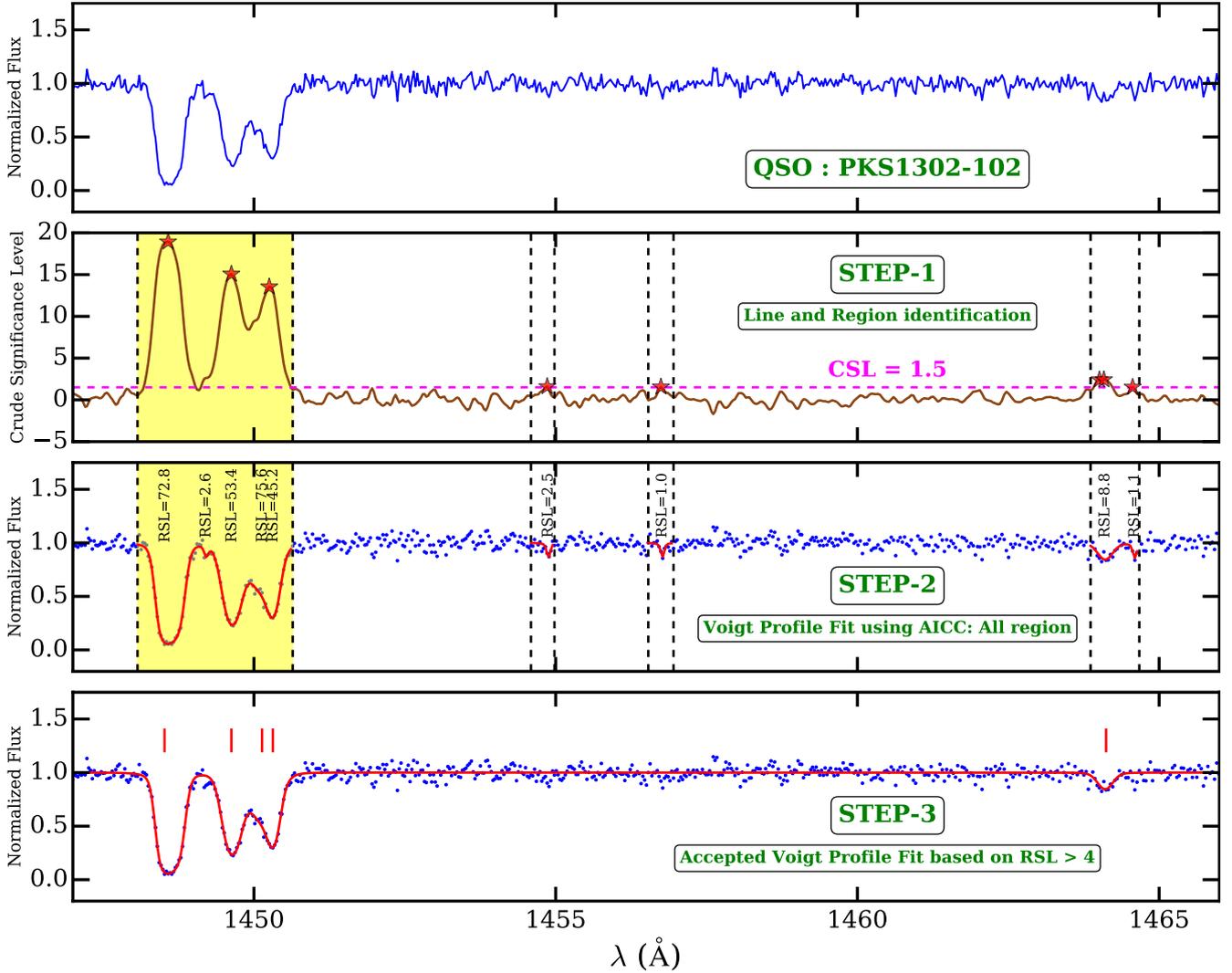
{Illustration of different steps in the automatic Voigt profile fitting procedure used in \codename. \textit{Top} panel shows a portion of the observed {\sc hst-cos} spectrum along the sightline towards QSO PKS1302-102. \textit{Second} panel from top shows the estimation of crude significance level (CSL) using the Eq.\ref{eq:csl} (for $b=20$ \kmps). All the identified peaks with ${\rm CSL} \geq 1.5$ (magenta dashed line) are shown by red stars. The identified regions enclosing the peaks are shown by black dashed vertical lines.
Overlapping regions are merged accordingly to fit blended lines simultaneously (see yellow shaded region). All the identified regions are fitted with Voigt profile as shown in the third panel from top. The number of components used to fit the region is decided using AICC and demanding $\chi^2_{\rm dof} \sim 1$ (see \S\ref{sec:method}). Rigorous significance level (RSL) for each fitted line is calculated using Eq.\ref{eq:rsl}. \textit{Bottom} panel shows the accepted fit with the ${\rm RSL} \geq 4$. 
}{\label{fig:vpfit-steps}}
\item \textbf{Voigt profile fitting : }
In this step we fit each identified region by multiple Voigt profiles.  Voigt profile, convolution of Gaussian with Lorentzian, is a real part of the ``Faddeeva function'' $w(z)$ \citep{armstrong1967},
\begin{equation}\label{eq:Voigt-fit}
\begin{aligned}
w(z)	&=	e^{-z^2} \: {\rm erfc}(-iz) \\ 
w(x + iy)	&=	V(x,y) + i L(x,y) 
\end{aligned}
\end{equation}
where ${\rm erfc(-iz)}$ is the error function, $V(x,y)$ is Voigt profile and $L(x,y)$ is imaginary part of Faddeeva function. We used \textit{wofz}\footnote{\url{http://ab-initio.mit.edu/wiki/index.php/Faddeeva_Package}} function in \textit{python's} \textit{scipy} package to compute Voigt profile. We convolve this Voigt profile with the appropriate {\sc hst-cos lsf} before performing $\chi^2$ minimization\footnote{We assumed that the observed fluxes in different pixels are uncorrelated. {\sc hst-cos lsf} is not a Gaussian (\url{http://www.stsci.edu/hst/cos/performance/spectral_resolution/}). This function is slightly asymmetric around the center and has extended wings.}. For $\chi^2$ minimization, we used \textit{leastsq}\footnote{\url{https://docs.scipy.org/doc/scipy-0.18.1/reference/generated/scipy.optimize.leastsq.html}} function  in \textit{python's} \textit{scipy} package. If $F_{\rm obs}$, $\sigma_{\rm obs}$, $F_{\rm fit}(\lambda_{\rm c},b,{\rm N_{HI}})$ are observed flux, error in the observed flux and fitted Ly-$\alpha$ flux respectively then we minimize the function  $(F_{\rm obs} - F_{\rm fit}) / \sigma^2_{\rm obs}$ in \textit{leastsq} routine. The fit parameters are allowed to vary over the range $5 \leq b \: ({\rm km \: s^{-1}}) \leq 150$, $10 \leq {\log({\rm N_{HI}} / {\rm cm^{-2}})} \leq 16.5$ and $\lambda_{\rm c}$ bounds are set by the wavelength of the region. We set initial guess values for lines in a given region by fitting individual line in that region with a Gaussian. Note that in the previous step each identified line is enclosed in two CSL minima hence we can fit each line separately. However initial guess values for any additional line required by the information criteria (explained below) are set randomly in the region.

A criteria based on information theory, Akaike Information Criteria with Correction (AICC) \citep{akaike1974,liddle2007,king2011} is used to assess the optimum number of Voigt profile components required for an acceptable fit. If $p$ is the number of parameters in a model used to fit the data with $n$ pixels, then the AICC is given by,
\begin{equation}\label{eq:aicc}
{\rm AICC} = \chi^2 + \frac{2 \:p \: n}{(n-p-1)} \;\;\;\; .
\end{equation}
The first term on right-hand side is a measure of loss of information while describing the data with a model.
The second term in right-hand side quantifies the complexity of the model. Thus AICC incorporates the trade-off between loss of information and complexity of the model.
We assign a model to be the best fit model over the previously assigned best fit model if AICC is lower by at least 5 \citep{jeffreys1961}. Since only the relative difference in the AICC values is important, according to \citet{jeffreys1961} $\Delta$AICC = 5 is considered as the strong evidence against the weaker model. Fig. \ref{fig:aicc-illustration} illustrates our method of choosing a best fit model. Black points in left-hand panel of the Fig. \ref{fig:aicc-illustration} are data points which we want to fit by Voigt profile model. We fitted the data with different number of (say $N_{\rm Voigt} = 1,2,3,4,5$) Voigt profiles. The resulting AICC and $\chi^2$ for each model as a function of $N_{\rm Voigt}$ is shown by star and circle respectively in the right-hand panel of the Fig. \ref{fig:aicc-illustration}.  
\InputFigCombine{AICC_illustartion.pdf}{175}
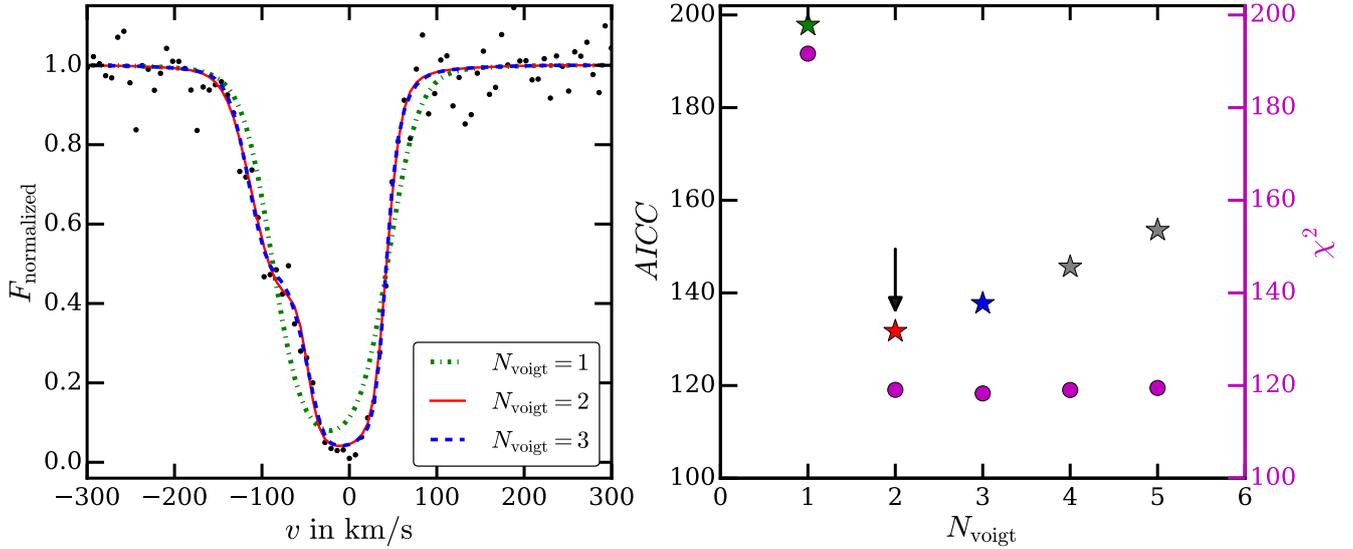
{\textit{Left-hand} panel shows the three different Voigt profile fits with $N_{\rm Voigt}=1,2$ and $3$ (green dot-dashed, red continuous and blue dashed lines respectively) fitted to the observed data (black circle). The spectrum is shown in the velocity scale defined with respect to the redshift of the strongest line center. \textit{Right-hand} panel shows the corresponding variation of AICC (stars) and $\chi^2$ (magenta circles) for 5 different models. For legibility fits with $N_{\rm Voigt} = 4,5$ (gray star points) are not shown in left-hand panel. For $N_{\rm Voigt} > 2$, the $\chi^2$ remains constant whereas AICC increases due to the second term on right-hand side of Eq.\ref{eq:aicc}. The best fit model corresponds to the minimum AICC (where $\chi^2_{\rm dof} \sim 1$ is also achieved) i.e., $N_{\rm Voigt} = 2$ shown by black arrow in right-hand panel and red solid line in left-hand panel.}{\label{fig:aicc-illustration}}
For $N_{\rm Voigt}=1$ the model is less complex but the $\chi^2$ ($\chi^2_{\rm dof} \sim 1.5$) between model and data is large (see green curve in left-hand panel) resulting in a larger AICC value\footnote{The reduced $\chi^2$ is given by, $$\chi^2_{\rm  dof} = \frac{\chi^2}{n-p}$$ where $n=134$ is number of pixels in the given region (left-hand panel of Fig. \ref{fig:aicc-illustration}), $p=3\times N_{\rm Voigt}$ is number of free parameters where factor 3 accounts for the number of free parameters in each Voigt component ($\lambda_{\rm c}$, ${\rm N_{HI}}$ and $b$).}. Whereas, for $N_{\rm Voigt} > 2$, the $\chi^2$ ($\chi^2_{\rm dof} \sim 1.0$) is small (see blue curve in left-hand panel) but with increasing $N_{\rm Voigt}$ the complexity of the model increases and hence AICC also increases. It is interesting to note that the $\chi^2$ remains nearly constant for $N_{\rm Voigt} \geq 2$ whereas AICC systematically increases for $N_{\rm Voigt} > 2$. A  model simply based on $\chi^2$ minimization, thus would be degenerate for $N_{\rm Voigt} \geq 2$. The minimum AICC occurs for $N_{\rm Voigt} = 2$ (black arrow showing red star in right-hand panel) which shows trade-off between goodness-of-fit ($\chi^2_{\rm dof} \sim 0.9$) and complexity of the model. The corresponding best fit model is shown by red curve in left-hand panel. Thus for minimum AICC, $\chi^2_{\rm dof}$ is also close to $1$ and hence we chose it to be the best fit model.
In the third panel from top of Fig. \ref{fig:vpfit-steps}, we show the results of fitting each region with as many Voigt components as necessary for the minimum AICC and $\chi^2_{\rm dof} \sim 1$. 
\item \textbf{Significance level of fitted lines :} Initially we fitted the lines that are identified using a simple approximation of ``Crude Significance Level''. However, we used a ``Rigorous Significance Level'' (hereafter RSL) formula \citep{keeney2012} to include the lines in the final line catalog as given below,
\begin{equation}\label{eq:rsl}
{\rm RSL} = {\rm (SNR)_1} \frac{W_{\lambda}}{\Delta \lambda} \: \frac{\eta(x)}{x} \: f_c(x,\lambda,b) \;\;\; .
\end{equation}
where, $x$ is width (in pixels) of discrete region over which equivalent width $W_{\lambda}$ is calculated, $\Delta \lambda = \lambda / x$, $\lambda$ is width (in ${\rm \AA}$) of the discrete region, $f_c(x,\lambda,b)$ is the fractional area of the {\sc hst-cos lsf} contained within the region of integration, $\eta(x) = {\rm (SNR)}_x / {\rm (SNR)_1}$ takes care of the fact that noise property may not be purely Poissonian, ${\rm (SNR)_1}$ is signal to noise ratio per pixel, ${\rm (SNR)}_x$ signal to noise ratio average over discrete region containing $x$ pixels. We used the parametric form of $\eta(x), f_c(x,\lambda,b)$ given by \citet[][their Eq. 4, Eq. 7 to Eq. 11 with parameters given in Table.1 for the coadded data]{keeney2012}. 

To avoid the spurious detection, we retain only feature measured with ${\rm RSL} > 4$ in the final line catalog. Other features are excluded from further analysis. Using this criteria, we find that the number of identified lines with \logNHI $\geq 12.4$ to be fitted by \codename (1277 H~{\sc i} Ly-$\alpha$ lines) are similar to those of \citetalias{danforth2016} (1280 H~{\sc i} Ly-$\alpha$ lines)\footnote{The total number of identified lines above completeness limit (i.e., \logNHI $\geq$ 13.6) in \codename and \citetalias{danforth2016} line catalog is 533 and 522 respectively.}.
In the third panel from top of Fig. \ref{fig:vpfit-steps}, we show the RSL for each fitted component above the line.  
Bottom panel of Fig. \ref{fig:vpfit-steps} shows that the final accepted Voigt profile fit that contains only those components which have ${\rm RSL} > 4$. 

The comparison of Voigt profile fit of \codename with that of \citetalias{danforth2016} method is illustrated in Fig. \ref{fig:comparison-spectra-danforth}. The fit from \codename and \citetalias{danforth2016} method is shown by solid red and dashed blue line respectively. The line centers of components fitted by \codename and \citetalias{danforth2016} method are shown by solid red and dashed blue vertical ticks respectively.  For \codename the reduced $\chi^2$ is small as compared to that for the components obtained by \citetalias{danforth2016}. The top row shows the example where \codename fit (in terms of number of component and the values of fitted parameter along with the errorbar) matches well with \citetalias{danforth2016} fit. In most situations the fitted parameters from \codename matched well with those from \citetalias{danforth2016} within errors. In few cases \codename fitted data better than \citetalias{danforth2016} method (bottom row of Fig. \ref{fig:comparison-spectra-danforth}) in terms of reduced $\chi^2$. We fitted all the observed spectra using \codename and form a line catalog ``\codename catalog''  (see Appendix for details). It should be noted that unlike \citetalias{danforth2016}, \codename does not fit higher order Lyman series lines (e.g. Ly-$\beta$, Ly-$\gamma$) simultaneously for an accurate measurement of \logNHI in the case of saturated Ly-$\alpha$ lines. However, we show in the next section that the differences in CDDF and line width distributions from ``\codename catalog'' and ``\citetalias{danforth2016} catalog'' are very small.

\InputFigCombine{Comparison.pdf}{175}
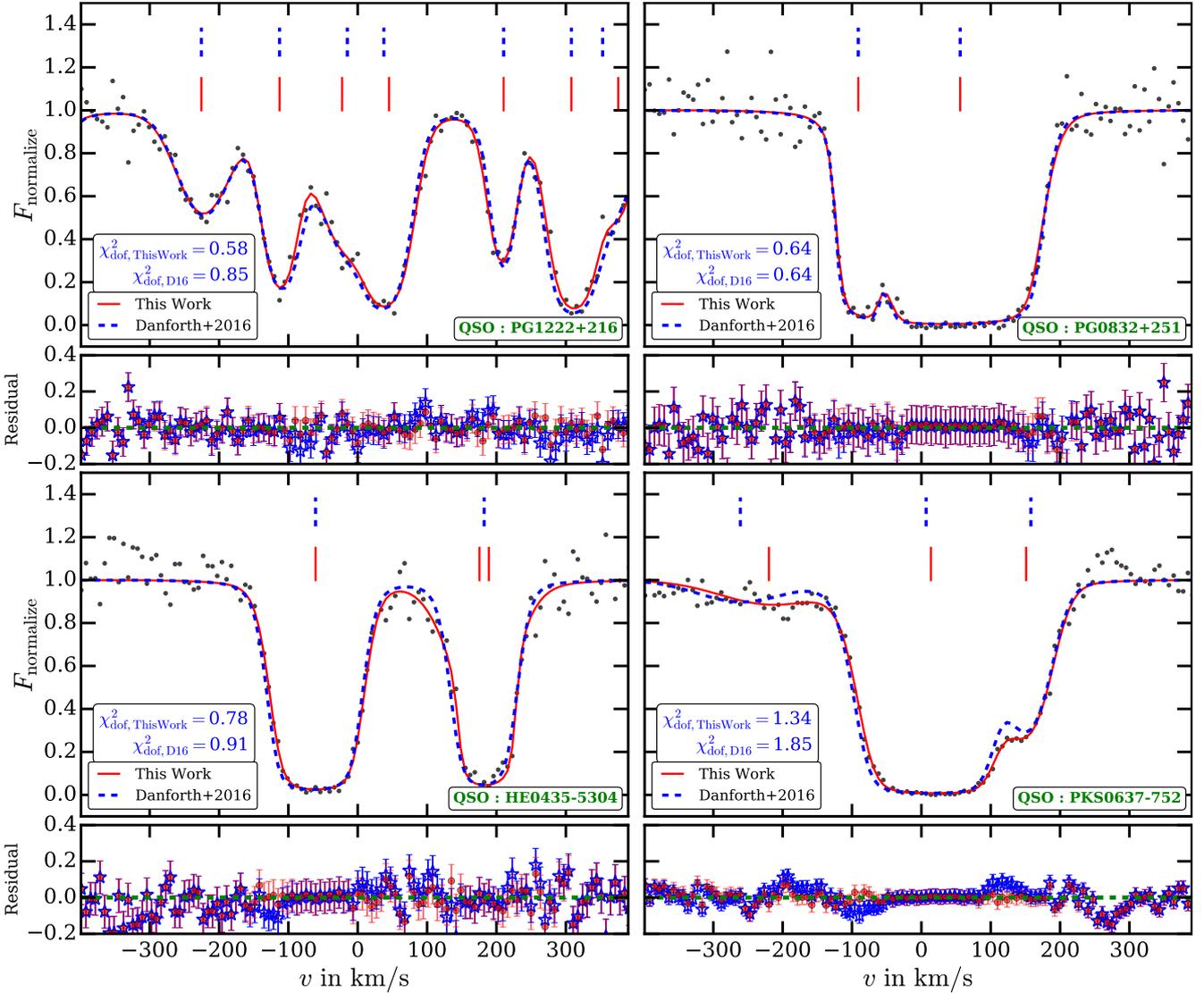
{Comparison of Voigt profiles fitted using our procedure with those of \citetalias{danforth2016} for four different regions in our sample. Black filled circles, the solid red line and the dashed blue line are observed data points, the best fit profile from \codename and from \citetalias{danforth2016} respectively. The spectra are shown in the velocity scale defined with respect to the redshift of the strongest component. Blue dashed and red continuous vertical ticks show the location of identified components by \citetalias{danforth2016} and \codename respectively. The residual between observed data and fitting from \citetalias{danforth2016} (open blue stars) and \codename (red filled circles) model are shown in the corresponding lower panel. In majority of cases ($\sim 89$ percent, like upper row panels) our parameters within $1\sigma$ errors match with those from \citetalias{danforth2016}. However, for some cases our fit to the data using AICC (i.e., using criteria $\Delta$AICC$\geq$ 5, see text for details) is found to be better (lower row panels). In all four cases shown above our $\chi^{2}_{\rm dof}$ is better than the corresponding from \citetalias{danforth2016}.
}{\label{fig:comparison-spectra-danforth}}

\subsection{Column density distribution function (CDDF):} Column density distribution function, $f({\rm N_{HI}},z)$, describes the number of absorption lines in the column density range \logNHI and \logNHI $+$ \dlogNHI and in the redshift range $z$ to $z+dz$. For a singular isothermal density profile of Ly-$\alpha$ absorbers, the H~{\sc i} photoionization rate \GHI can be inferred from H~{\sc i} CDDF as \citep{schaye2001,shull2012b},
\begin{equation}\label{eq:column-density}
f({\rm N_{HI}},z) = \frac{\partial^2 N}{\partial z \; \partial {\rm \log(N_{HI}})} \propto \Gamma^{-1/2}_{\rm HI} \;\;\; .
\end{equation}
We take into account the completeness of the sample while calculating the redshift path length as a function of \logNHI. Following \citetalias{danforth2016}, we calculate the CDDF in $13$ \logNHI bins with centers at $12.5, 12.7, \cdots, 14.7,14.9$ and width \dlogNHI $= 0.2$.
\InputFigCombine{z_path_illustration.pdf}{175}
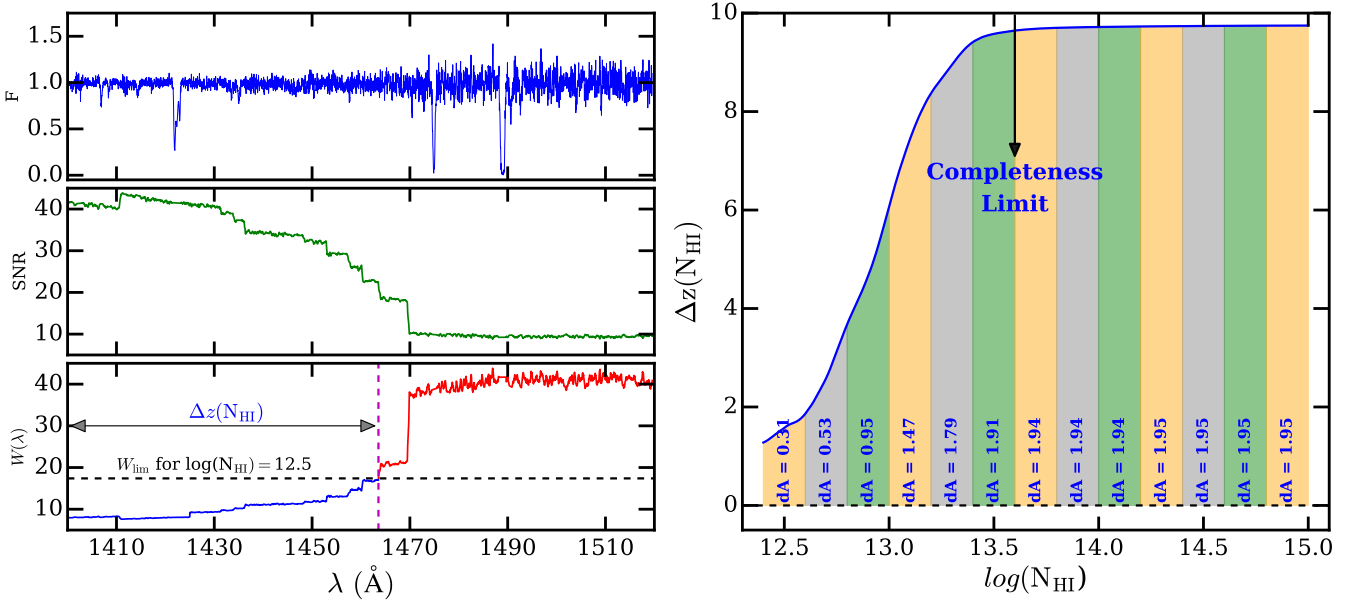
{\textit{Left-hand} panel illustrates our redshift path length calculation for sightline towards QSO H1821+643. Top, middle and lower left-hand panels show the flux, SNR per pixel and equivalent width vector respectively. Equivalent width vector $W(\lambda)$ is calculated for RSL = 4 in Eq.\ref{eq:rsl}. The limiting equivalent width ($W_{\rm lim}$), estimated from the curve of growth, corresponding to $\log({\rm N_{HI}}) = 12.5$ is shown by horizontal black dashed line in the bottom panel.  The redshift path length, $\Delta z ({\rm N_{HI}} = 10^{12.5} \; {\rm cm^{-2}})$, for this sightline is the redshift covered by region $W(\lambda) \leq W_{\rm lim}$. 
The total redshift path length is sum of the $\Delta z$ measured along all QSO sightlines. \textit{Right-hand} panel shows (blue curve) the total redshift path length as a function of \logNHI (known as sensitivity curve).
The completeness limit for the sample is \logNHI $=13.6$ (shown by blue arrow). The fractional area in a given \logNHI bin, $dA = dz \; d\log({\rm N_{HI}})$, is area under the blue curve in the corresponding  \logNHI bin (shown by blue text) that is used in CDDF calculation.}{\label{fig:redshift-path-length-illustration}}

Left-hand panel in Fig. \ref{fig:redshift-path-length-illustration} illustrates the procedure we adopt for calculating the redshift path length. The top, middle and lower sub-panels show the flux $F$, SNR per pixel and equivalent width vector $W(\lambda)$ respectively for a sightline towards QSO H1821+643. To calculate the equivalent width vector $W(\lambda)$, Eq.\ref{eq:rsl} is rearranged and solved for $W(\lambda)$ by taking RSL=4 and $b=17 \; {\rm km \: s^{-1}}$ (corresponds velocity resolution) for each pixel. Next we calculate limiting equivalent width $W_{\rm lim}$ from curve of growth ($b =17 \; {\rm km \: s^{-1}}$) for different values of \logNHI. As an example we show $W_{\rm lim}$ for \logNHI $ = 12.5$ by black dashed horizontal line in bottom panel of Fig. \ref{fig:redshift-path-length-illustration}. The redshift path length $\Delta z({\rm N_{HI}} = 10^{12.5} \; {\rm cm^{-2}})$ for this sightline (shown by blue curve in bottom panel) is the sum of redshift range for which $W(\lambda) \leq W_{\rm lim}$. The total redshift path length $\Delta z({\rm N_{HI}})$ covered in the observed sample is a sum of all the redshift path length in individual sightlines. The total redshift path length $\Delta z({\rm N_{HI}})$ is then plotted as a function of \logNHI (`Sensitivity curve') as shown in right-hand panel of Fig. \ref{fig:redshift-path-length-illustration}. The completeness limit for the sample is \logNHI $=13.6$ (shown by blue arrow) i.e., the lines with \logNHI $\geq13.6$  are always detectable over the entire observed wavelength range for the full sample. The fractional area $dA = dz \times$\dlogNHI in Eq.\ref{eq:column-density} is calculated by integrating the sensitivity curve in the corresponding \dlogNHI bin. We shall refer to the CDDF obtained for our fitted parameters (over the redshift range $0.075 \leq z \leq 0.45$) in this way as ``\codename CDDF''.

In the left-hand panel of Fig. \ref{fig:comparison-cdd-danforth}, we compared the \codename CDDF with CDDF given in Table.5 of \citetalias{danforth2016} (also \citetalias{shull2015}) and CDDF calculated from \citetalias{danforth2016} line catalog. The CDDF given in Table.5 of \citetalias{danforth2016} is calculated from $2256$ H~{\sc i} absorbers in the redshift range $0 \leq z < 0.75$ whereas CDDF calculated from \citetalias{danforth2016} line catalog contains $1280$ H~{\sc i} absorbers in the redshift range $0.075 \leq z < 0.45$. We used our redshift path length estimation for CDDF calculation from \citetalias{danforth2016} line catalog.
The \codename and the \citetalias{danforth2016} line catalog CDDFs are consistent with each other within $1\sigma$ except at high and low \logNHI bins. The median \logNHI from \codename and \citetalias{danforth2016} is $13.39 \pm 0.61$ and $13.38 \pm 0.63$ respectively. The two sample KS test $p$-value between the \logNHI distribution of \codename and \citetalias{danforth2016} line catalog is 0.83.
The consistency between \codename CDDF and \citetalias{danforth2016} line catalog CDDF suggests that the number of components identified by \codename in different \dlogNHI bins are similar to those from \citetalias{danforth2016} line catalog. Whereas, the agreement between \codename CDDF and CDDF from \citetalias{danforth2016} paper indicates that our redshift path calculation is consistent with that from \citetalias{danforth2016} paper.

We notice occasional component differences when the Ly-$\alpha$ line is heavily saturated between \citetalias{danforth2016} and our \codename fits. Unlike \citetalias{danforth2016}, \codename does not include simultaneous fitting of Ly-$\alpha$ line and higher order Lyman series lines (such as Ly-$\beta$, Ly-$\gamma$). This could be the reason for minor mismatch of CDDF in the bins \logNHI $=13.5-13.7$ and \logNHI $=14.7-14.9$. Thus minor differences one notices in high and intermediate \logNHI bins in observed CDDF can be attributed to the differences in the multi-component fitting procedure in particular to the way the total number of components fitted to a given identified absorption region. However, there is an overall good agreement between CDDF derived by \citetalias{danforth2016} and \codename.

In the right-hand panel of Fig. \ref{fig:comparison-cdd-danforth}, we compared the doppler-$b$ parameter distribution from \codename catalog (red curve with shaded region) and \citetalias{danforth2016} line catalog (black dashed line with error bars)\footnote{In left-hand panel of Fig. \ref{fig:comparison-cdd-danforth}, we compare our CDDF with that from \citetalias{danforth2016} line catalog ($0.075 \leq z \leq 0.45$) and \citetalias{danforth2016} paper ($0 \leq z \leq 0.75$). However, in right-hand panel of Fig. \ref{fig:comparison-cdd-danforth}, we compare our $b$ parameter distribution (i.e. in the redshift range $0.075 \leq z \leq 0.45$) with that from \citetalias{danforth2016} catalog only as the $b$ parameter distribution is not available in \citetalias{danforth2016} paper.}. In both distributions the errors are assumed to be poisson distributed.
The median value of $b$ parameter from \codename and \citetalias{danforth2016} is  $32.9 \pm 20.8$ \kmps and $33.9 \pm 18.3$ \kmps respectively. The two sample KS test $p$-value between the $b$ parameter distribution of \codename and \citetalias{danforth2016} line catalog is 0.41. Thus the two distributions are consistent with each other validating the consistency between \codename and \citetalias{danforth2016} line fitting methods.
\end{enumerate}
\InputFigCombine{CDD_BPD_Danforth.pdf}{175}
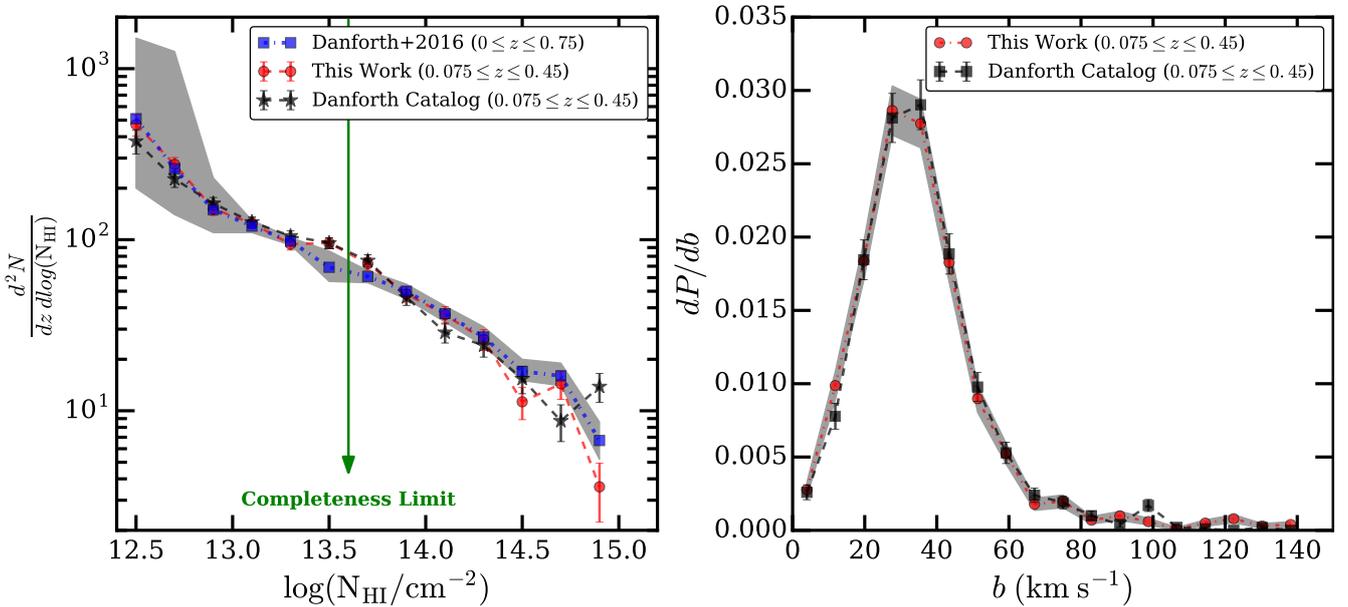
{Left-hand panel shows comparison of CDDF from \codename (red circle, $0.075 \leq z \leq 0.45$, 1277 H~{\sc i} Ly-$\alpha$ lines), \citetalias{danforth2016} line catalog (black stars, for $0.075 \leq z \leq 0.45$, 1280 H~{\sc i} Ly-$\alpha$ lines) and Table.5 in \citetalias{danforth2016} (blue square curve with gray shaded region, for $0 \leq z \leq 0.75$, 2256 H~{\sc i} Ly-$\alpha$ lines). The two sample KS test $p$-value between \codename and \citetalias{danforth2016} line catalog for \logNHI distribution is 0.83. Thus within errors the CDDF from the two methods are consistent with each other. At high column densities the differences arises due to differences in the fitting procedure (multi-component fitting using AICC). Right-hand panel shows the $b$ parameter distribution from \codename (red curve for $0.075 \leq z \leq 0.45$) and \citetalias{danforth2016} line catalog (black curve with shaded region for $0.075 \leq z \leq 0.45$). The two sample KS test $p$-value between \codename and \citetalias{danforth2016} line catalog for $b$ parameter distribution is 0.41. Thus the $b$ parameter distribution from \codename is in good agreement with that of \citetalias{danforth2016} line catalog validating our procedure. In both panels the error bars shown are computed assuming the Poission distribution. }{\label{fig:comparison-cdd-danforth}}
\vspace{-3mm}
\section{Simulation}
\label{sec:simulation}
In this section we discuss the simulations used to generate the Ly-$\alpha$ forest that will be subsequently used to constrain \GTW in \S\ref{sec:results}.
We used publicly available smoothed particle hydrodynamic code {\sc gadget-2}\footnote{\url{http://wwwmpa.mpa-garching.mpg.de/gadget/}} \citep{springel2005} to generate density and velocity field in a periodic box of size $50 \; h^{-1}$ \cmpc. The initial conditions were generated at $z=99$ by using the publicly available {\sc 2lpt}\footnote{\url{http://cosmo.nyu.edu/roman/2LPT/}} \citep{2lpt2012} initial conditions code.
Radiative heating and cooling processes are not incorporated in default version of {\sc gadget-2}. We used our module ``Code for Ionization and Temperature Evolution'' (hereafter {\sc cite}) to evolve the temperature of the IGM in the post-processing step of the {\sc gadget-2}. We refer reader to \citetalias{gaikwad2016} where {\sc cite} is discussed in detail. 

In \citetalias{gaikwad2016}, we showed the consistency of our method of evolving IGM temperature in {\sc gadget-2}+{\sc cite} with other low-$z$ simulations in the past \citep{dave2010,smith2011,tepper2012} using 3 metrics (i) the thermal history parameters at $z < 0.5$, (ii) distribution of baryons in different regions of the phase diagram and (iii) column density (${\rm N_{HI}}$) vs overdensity ($\Delta$) relationship. 
We used model $T15-\gamma1.3$ given in Table 3 of \citetalias{gaikwad2016} to calculate column density distribution function. In this model we evolve the equation of state of the IGM from $T_0 = 15000$ K and $\gamma=1.3$ at $z=2.1$ to $z=0$. The final thermal history parameters at low redshift for this model are $(z,T_0,\gamma) \equiv $ $(0.4,5220,1.51)$, $(0.3,4902,1.53)$, $(0.2,4583,1.54)$, $(0.1,4245,1.55)$

Once we have calculated temperature of the {\sc gadget-2} particles we generate the Ly-$\alpha$ forest by shooting random lines of sight through simulation box \citep[][\citetalias{gaikwad2016}]{trc2001,hamsa2015}. To account for the possible redshift evolution of the structures, we generated Ly-$\alpha$ forest from the simulation box at $\overline{z} = 0.1, 0.2, 0.3$ an $0.4$ for comparison with observations. The basic steps are as follows,
\begin{itemize}
\item Using {\sc sph} smoothing kernel, the density, velocity and temperature are calculated on the grids along the sightline.
\item Assuming photoionization equilibrium with ultra-violet background (hereafter UVB), the number density of neutral hydrogen ($n_{\rm HI}$) is calculated along the sightline.
\item We treat H~{\sc i} photoionization rate (\GHI) as a free parameter in our analysis. For a given value of \GTW, Ly-$\alpha$ optical depth $\tau$ is calculated at each pixel from the $n_{\rm HI}$ field and by properly accounting for peculiar velocity, thermal width and natural width effects on the line profile.The transmitted flux is then given by $F=\exp(-\tau)$. 
\item The velocity separation of pixels in the simulated spectra is $\sim 5 \; {\rm km \: s^{-1}}$ whereas observed spectra are at $\sim 17 \; {\rm km \: s^{-1}}$ velocity resolution. We linearly interpolate the simulated spectra to match with observed velocity resolution.
\item The line spread function of {\sc hst-cos} is skewed and has extended wings. We account for this effect by convolving the simulated spectra with {\sc hst-cos lsf}. Finally we add the Gaussian random noise to the simulated spectra with SNR same as the observed spectra.  
\end{itemize}

We generated $N \times N_{\rm spec}$ number of Ly-$\alpha$ forest spectra at redshift $z$ to account for the variance in the model CDDF \citep[for similar method see][\citetalias{gaikwad2016}]{rollinde2013}. Here, $N_{\rm spec}$ is number of observed spectra at the redshift of interest $z$ and $N=100$ is number of mock samples. It should be noted that the dynamical impact of diffuse IGM pressure is not modeled self-consistently in {\sc cite}. However, these effects are not important for low resolution ($\sim 50$ \ckpc) simulations presented in this work (see \citetalias{gaikwad2016} for details).
Our simulations do not contain any higher order (i.e., Ly-$\beta$, Ly-$\gamma$ and onward) Lyman series lines for most of our analysis. However, we consider the contamination by higher order Lyman series (in particular Ly-$\beta$ from $z=0.6$) lines with the Ly-$\alpha$ line for the highest-$z$ bin i.e., $z=0.4$ (see \citetalias{gaikwad2016}).
\section{Results}
\label{sec:results}
In this section we decomposed Ly-$\alpha$ forest spectra generated from {\sc gadget-2} + {\sc cite} simulations into multiple Voigt profiles using \codename. We formed a line catalog for each mock sample and obtained three distributions (i) $b$ parameter distribution, (ii) $b$ vs \logNHI distribution and (iii) CDDF. These distributions calculated from different mock samples are used to estimate the errors. We compare these distributions from simulations with those from observation. In particular, we used CDDF to constrain the \GTW and its evolution in four different redshift bins.

\subsection{$b$ parameter distribution}
\label{sec:b-parameter-distribution}
The $b$ parameter distribution calculated from Voigt profile fitting  is sensitive to thermal history, the energy injected by various astrophysical processes in the form of heat and unknown turbulent motions in the IGM \citep{schaye1999,schaye2000,mcdonald2001,dave2001a}.
Recently it has been found that the $b$ parameters obtained from various hydrodynamical simulations are typically smaller compared to those from the observations at low-$z$  \citep{viel2016} though the thermal history parameters self-consistently obtained in these simulations agree well with each other. 
The $b$ parameters from our simulation ({\sc gadget-2} + {\sc cite}) are also found to be significantly smaller than the observed $b$ parameters (see left-hand panel of Fig. \ref{fig:b-pdf-distribution}).
Thus to match the $b$ parameter distribution, additional thermally and/or non-thermally broadened $b$ parameter is required.

We found that the observed $b$ parameter distribution can be matched with simulation by increasing the temperature of each pixel along sightline by a factor of $3$. This increase in temperature would correspond to injection of the energy in the form of heat into the IGM. 
Increasing the gas temperature can lead to two effects (i) reducing the recombination rate coefficient as it scales as $\sim T^{-0.7}$ and (ii) introduce additional ionization due to collisions in the high density gas. By artificially enhancing the heating rate by more than a factor of $3$, \citet{viel2016} have recently shown the simulated $b$ distribution can be made consistent with the observed ones. However such a model tends to suggest lower \GHI compared to simulations without additional heating \citep[see Table 1 in][]{viel2016}. As we will show in \S\ref{subsec:gamma-12-implications}, a lower value of \GHI would imply that the number of ionizing photons in the IGM is less than that expected from only QSOs. Hence we rather focus on a scenario where the additional contribution to the line broadening arises from non-thermal motions. Here we explore this possibility by introducing turbulent motions that can simultaneously explain $b$ parameter distributions and $b$ vs \logNHI distributions as well.

In this work we incorporate additional broadening by adding a \textit{non-thermal} (micro-turbulence) component $b_{\rm turb}$ to thermal $b$ parameter in quadrature ($b^2 = b^2_{\rm thermal} + b^2_{\rm turb}$) to mimic micro-turbulence that is missing in our simulation and see its effect on the derived \GTW constraints. We refer to the non-thermal contribution to the $b$ parameters as `micro-turbulence', $b_{\rm turb}$.  In general, this `micro-turbulence' is due to any physical phenomenon affecting the width of the absorption line that is not captured properly in  our simulations e.g. various feedback processes and/or numerical effects\footnote{The physical phenomena occurring on scales below the resolution scale (i.e., below $\sim 50h^{-1}$ \ckpc) of the simulation box may affect the scales that are resolved \citep{springel2002}. These physical phenomena may not be captured properly in the simulation box.}.
Using this new $b$ parameter we compute the Ly-$\alpha$ optical depth and fit each absorption line using \codename to get $b$ and \logNHI. 
The \GTW is then constrained from the model with and without micro-turbulence.
We used two different models to quantify the micro-turbulence in simulation as explained below. 
\begin{enumerate}
\item Density dependent $b_{\rm turb}$: In order to match the observed line width distribution of  O~{\sc vi} absorbers, \citet{oppenheimer2009} added density dependent turbulence in their simulation. Following Eq. 5 and 6 given in \citet{oppenheimer2009}, we added (in quadrature) the $b_{\rm turb}({\rm n_H})$ in simulated spectra where ${\rm n_H}$ is hydrogen number density. The form of these equations is such that the contribution of the $b_{\rm turb}$ is appreciable only at high column densities i.e. \logNHI $> 13.5$ (see middle panel in Fig. \ref{fig:b-vs-NHI-distribution}). The $b$ parameter distribution for this case is shown in the middle panel of Fig. \ref{fig:b-pdf-distribution}. It is clear that the simulated $b$ values are smaller than the observed ones even in this case.
\item Gaussian random $b_{\rm turb}$: In this approach we generated $b_{\rm turb}$ from a Gaussian random variable with mean $\mu = 20$\kmps and standard deviation $\sigma=10$\kmps at each grid point along a sightline in the simulation box\footnote{We calculated $b$ vs \logNHI for $5$ different ($\mu, \sigma$) $\equiv$ (10,10),(20,5),(20,10),(20,15),(30,10) combinations. The $\chi^2$ is found to be minimum for $\mu = 20$\kmps and $\sigma=10$\kmps.}. These values are in agreement with the distribution of non-thermal broadening parameters \citep[see Fig. 24 in][]{tripp2008,muzahid2012} derived for the well-aligned O~{\sc vi} and H~{\sc i} absorbers.
From Fig. \ref{fig:b-pdf-distribution} (right-hand panel), we see that the agreement between the observed and model $b$ parameter distribution is better in the case of Gaussian distributed $b_{\rm turb}$ model than the other two models i.e., without any additional $b_{\rm turb}$ and density dependent $b_{\rm turb}$ models. The model results shown in Fig. \ref{fig:b-pdf-distribution} were based on simulations that use the best fitted redshift evolution of \GTW as given in \citetalias{gaikwad2016}. However, we also found that the $b$ parameter distribution depends weakly on the assumed evolution of \GHI.
\end{enumerate}

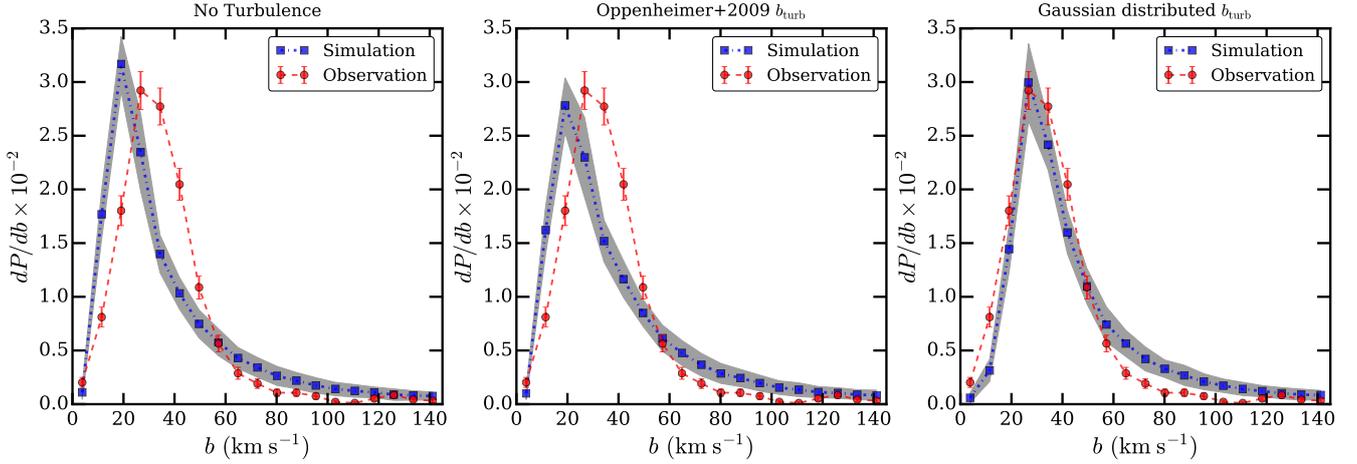
\InputFigCombine{b_pdf_1D.pdf}{175}{Comparison of $b$ parameter distribution (at $0.075 \leq z \leq 0.45$) from observations (red dashed line with errorbar) and simulations (blue dotted line with $1\sigma$ shaded region) for 3 cases (see \S\ref{sec:b-parameter-distribution}): (i) when $b_{\rm turb}$ is not added in the simulation (\textit{left-hand} panel), (ii) when density dependent $b_{\rm turb}$ at given $n_{\rm H}$ \citep{oppenheimer2009} is added in the simulation (\textit{middle panel}) and (iii) when Gaussian distributed $b_{\rm turb}$ is added in the simulation (\textit{right-hand} panel). The errorbars on model $b$ parameter distribution are calculated from mock sample whereas the errorbars on observed $b$ parameter distribution are calculated assuming Poisson statistics. The $b$ parameter distribution from models with Gaussian distributed $b_{\rm turb}$ qualitatively matches well with that from the observation.}{\label{fig:b-pdf-distribution}}

\subsection{$b$ vs \logNHI distribution}
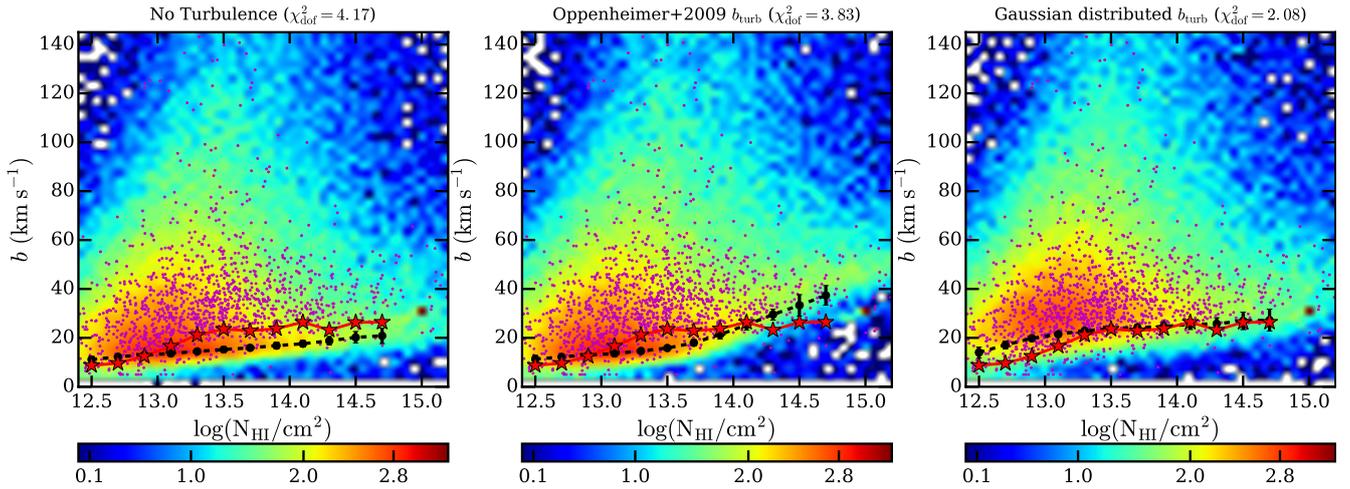
\InputFigCombine{b_vs_NHI_image.pdf}{175}{Comparison of $b$ vs ${\rm N_{HI}}$ distribution (at $0.075 \leq z \leq 0.45$) from observation (magenta points) and simulation (color-coded diagram) for 3 cases (see \S\ref{sec:b-parameter-distribution}) (i) when $b_{\rm turb}$ is not added in the simulation (\textit{left-hand} panel), (ii) when density dependent $b_{\rm turb}$ at given $n_{\rm H}$ \citep{oppenheimer2009} is added in the simulation (\textit{middle panel}) and (iii) when Gaussian distributed $b_{\rm turb}$ is added in the simulation (\textit{right-hand} panel). The color scheme indicates density of points from the simulation in logarithmic units. The red solid line and black dashed line shows the lower envelope for observed and model data points in both panels. The lower envelope is obtained by calculating $10^{\rm th}$ percentile of $b$ in \logNHI bins. The lower envelope matches well in the case where Gaussian distributed $b_{\rm turb}$ is added. We calculated the $\chi^2$ between model and observation by binning the data into 2D bins. These values are quoted on top of each panel. The $\chi^2_{\rm dof}$ is better for a model with Gaussian distributed $b_{\rm turb}$ ($\chi^2_{\rm dof} = 2.08$) than a model without turbulence ($\chi^2_{\rm dof} = 4.17$) and a model with density dependent $b_{\rm turb}$ ($\chi^2_{\rm dof} = 3.83$).}{\label{fig:b-vs-NHI-distribution}}

In this section we discuss the effect of adding $b_{\rm turb}$ on the $b$ vs \logNHI 2D distribution\footnote{We refer reader to \citet{webb1991,fernandez1996} for discussions on Voigt profile fitting procedure influencing this correlation.}. Fig. \ref{fig:b-vs-NHI-distribution} shows comparison of $b$ vs \logNHI distribution ($0.075 \leq z \leq 0.45$) from observations (shown by magenta points) with that from $3$ different models (i) without turbulence (left-hand panel), (ii) density dependent $b_{\rm turb}$ as suggested by \citet[][middle panel]{oppenheimer2009} and (iii) Gaussian distributed $b_{\rm turb}$ (right-hand panel).
The color scheme indicates the density of points in logarithmic units.
To assess the goodness-of-fit, we calculated the reduced $\chi^2$ for 2D distribution by binning the data along $b$ axis in 21 bins (with bin width 7.5 \kmps) and along \logNHI axis in 13 bins (with bin width 0.2). Let $P_{\rm sim,k}(i,j)$ be the value of 2D distribution of $k^{\rm th}$ mock sample in $i^{\rm th}$, $j^{\rm th}$ bin along \logNHI, $b$ axis respectively. The mean and standard deviation of 2D distribution from mock samples can be calculated as,
\begin{equation}\label{eq:mean-sigma-cdd}
\begin{aligned}
\overline{P}_{{\rm sim}}(i,j) &= \frac{1}{N} \sum \limits_{k=1}^{N} P_{{\rm sim},k}(i,j) \\ 
\sigma^2_{{\rm sim}}(i,j) &= \frac{1}{N-1} \sum \limits_{k=1}^{N}  [P_{{\rm sim},k}(i,j) - \overline{P}_{{\rm sim}}(i,j)]^2
\end{aligned}
\end{equation}
where $N=100$ is the number of mock samples. Let $P_{\rm obs}(i,j)$ be the value of observed 2D distribution in $i^{\rm th}$, $j^{\rm th}$ bin along \logNHI, $b$ axis respectively. Note that both distributions i.e., $P_{\rm obs}(i,j)$ and $P_{\rm sim,k}(i,j)$ are normalized. The reduced $\chi^2$ between the observed and model distribution is,
\begin{equation}
\begin{aligned}
\chi^2_{\rm dof} = \frac{1}{N_{\rm y} \times N_{\rm x} -1} \sum \limits_{i=1}^{N_{\rm x}} \sum \limits_{j=1}^{N_{\rm y}} \frac{[\overline{P}_{{\rm sim}}(i,j) - P_{\rm obs}(i,j)]^2}{\sigma^2_{{\rm sim}}(i,j)}
\end{aligned}
\end{equation}
where $N_{\rm x} = 13$ and $N_{\rm y}=21$ is number of bins along \logNHI and $b$ axis respectively.
From Fig. \ref{fig:b-vs-NHI-distribution} the $\chi^2_{\rm dof}$ for Gaussian distributed $b_{\rm turb}$ is better ($\sim 2.1$) than the model without turbulence ($\chi^2_{\rm dof} \sim 4.2$) and model with density dependent $b_{\rm turb}$ ($\chi^2_{\rm dof} \sim 3.8$).

Another way to asses the goodness of the assumed form for $b_{\rm turb}$ is to match the lower-envelope in $b$ vs \logNHI distribution. At $z  > 2$, the lower envelope is strongly correlated with thermal history parameters and has been used in the past to measure the effective equation of state of the IGM at high-$z$ \citep{schaye1999,schaye2000}. The red stars with solid line and black circles with dashed line in Fig. \ref{fig:b-vs-NHI-distribution} shows the lower envelope for observed and model $b$ vs \logNHI distribution respectively. The lower envelope is obtained by calculating the $10^{\rm th}$ percentile of $b$ values in each \logNHI bin \citep{garzilli2015a}. In model $b$ vs \logNHI distribution, the lower envelope is calculated  for all mock samples. The black circles with errorbars in Fig. \ref{fig:b-vs-NHI-distribution} represents mean and standard deviation of lower envelope from mock samples. In the case of models without turbulence we see that the lower envelope obtained for the observed data is systematically higher than that from simulation for \logNHI $\geq 13.2$. It is also clear from the middle panel that density dependent turbulence overproduce $b$ at \logNHI $ > 14.2$ and under produce in the range $13 \leq$ \logNHI $\leq 14$. It is evident from Fig. \ref{fig:b-vs-NHI-distribution} that the lower envelope in Gaussian distributed $b_{\rm turb}$ model (right-hand panel) matches well with the observed lower envelope at \logNHI $\geq 13.2$\footnote{The matching between observation and model with Gaussian distributed $b_{\rm turb}$ is also good when lower envelope is calculated using $5^{\rm th}$ and $20^{\rm th}$ percentile.}. At low column densities i.e.,  \logNHI $< 13.2$ the observed $b$ parameters tend to be smaller than what is predicted in this case. This suggests that the actual $b_{\rm turb}$ could be smaller at \logNHI $< 13.2$ compared to mean value we have assumed. Note at  these low \logNHI values Ly-$\alpha$ absorption are in the linear part of curve of growth and ${\rm N_{HI}}$ measurements are independent of $b$ parameter.

In summary the Gaussian distributed $b_{\rm turb}$ model matches well with observation for $b$ vs \logNHI distribution and $b$ parameter distribution. We reemphasize that this may not be the unique explanation for the additional broadening required in the simulation even-though it consistently reproduces the $b$ parameter distribution and $b$ vs \logNHI scatter. In the next section, we calculate CDDF and constrain \GTW from observation  by comparing model with Gaussian distributed $b_{\rm turb}$ and model without any $b_{\rm turb}$.

\subsection{\GHI Constraints using CDDF}
\label{subsec:Gamma-constraints}
\InputFigCombine{Constraint_Combine.pdf}{170}
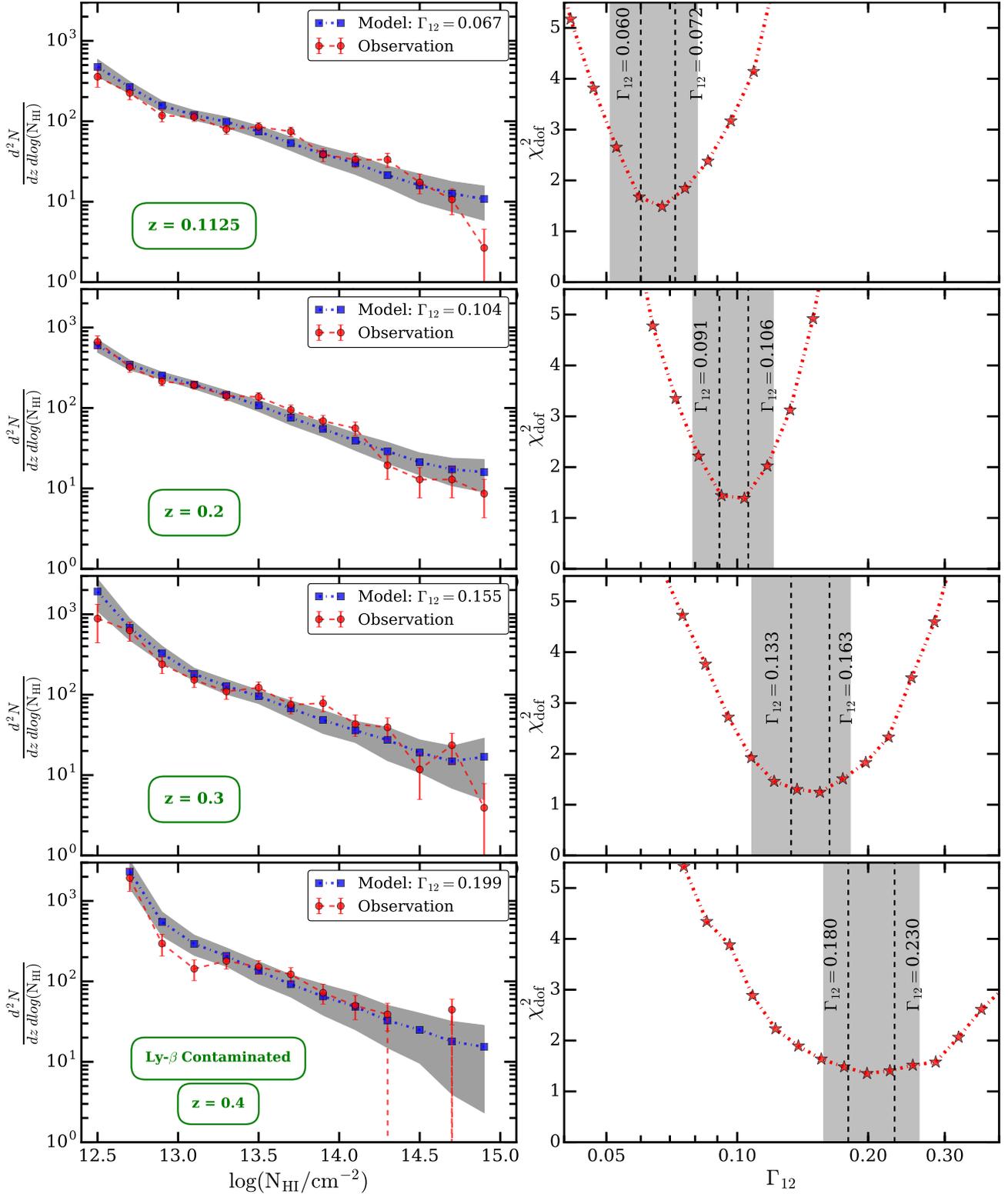
{Constraints on \GTW from CDDF in four different redshift bins. The \textit{left-hand} panels show the observed CDDF and CDDF from the simulation using the best fitted \GTW, i.e., \GTW corresponding to the minimum reduced $\chi^2$, in different redshift bins (given in green box). The gray shaded region shows $1\sigma$ errorbar on the model CDDF (calculated from mock samples). The errors on observed CDDF (red line with errorbar) are obtained assuming poisson distribution and are not considered for calculating  the reduced $\chi^2$. The \textit{right-hand} panels show the reduced $\chi^2$ as a function of the assumed \GTW. The black dashed vertical lines indicate $1\sigma$ constraints on \GTW around the \GTW where reduced $\chi^2$ is minimum. The shaded region indicates $1\sigma$ constraints on \GTW from flux PDF and flux PS given in \citetalias{gaikwad2016}. The $1\sigma$ constraints on \GTW from CDDF are well within those obtained using flux PDF and flux PS.  The simulated Ly-$\alpha$ forest at $z=0.4$ is contaminated by Ly-$\beta$ forest from $z \sim 0.6$ in the same wavelength range to account for possible contamination due to intervening H~{\sc i} absorbers (see \S \ref{sec:results} for more details).}{\label{fig:gamma-12-constraint}}

\InputFigCombine{Constraint_Combine_Turbulence.pdf}{170}
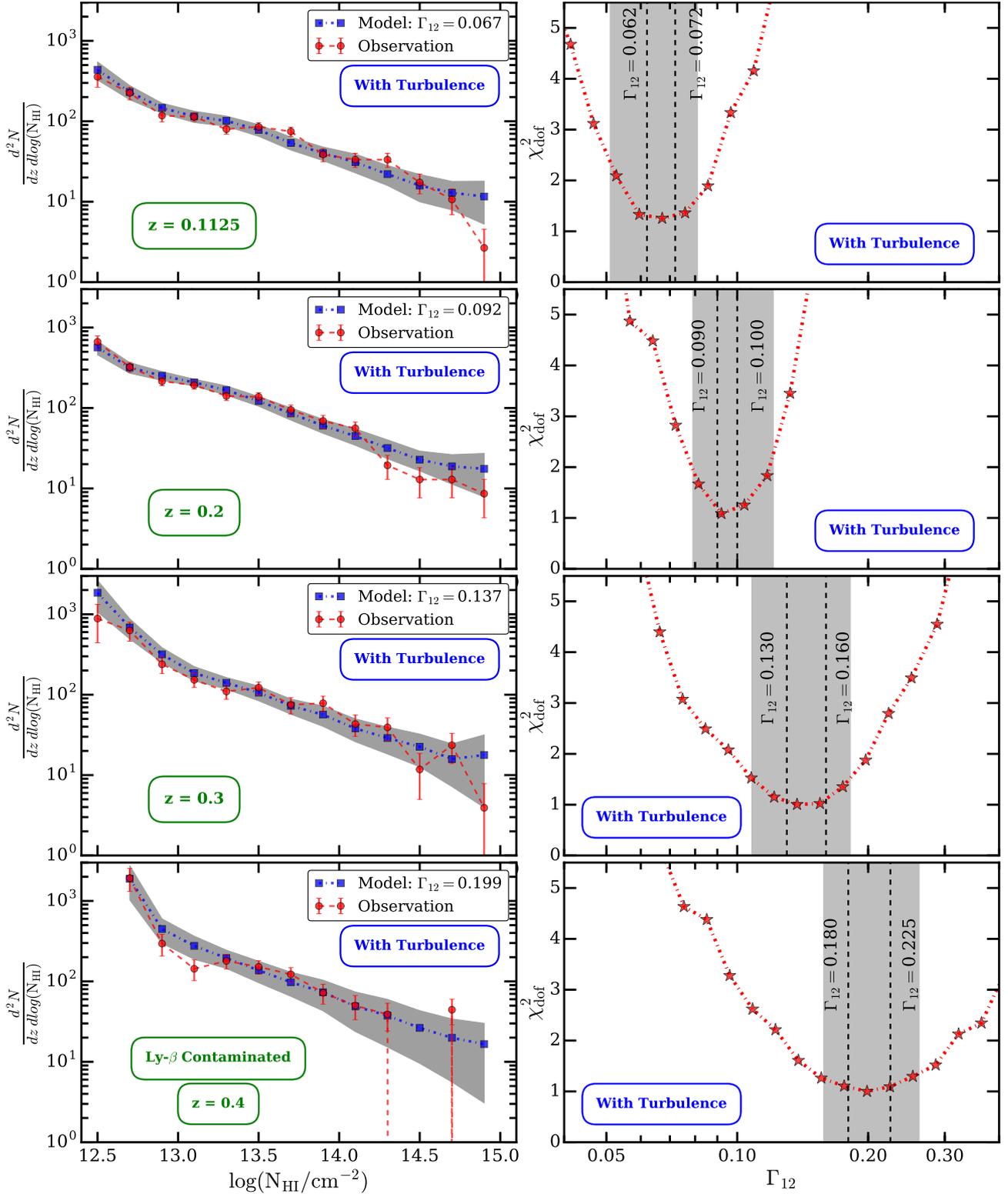
{Same as Fig. \ref{fig:gamma-12-constraint} except \GTW is constrained using a model with Gaussian distributed $b_{\rm turb}$ see \S \ref{sec:b-parameter-distribution}.}{\label{fig:gamma-12-constraint-turb}}

In this section we match  model CDDF with the observed CDDF to constrain \GTW in 4 redshift bins identified above.  The CDDF is calculated for each mock sample. \GTW is a free parameter in our analysis and the model CDDF depends on its value.  The model CDDF is binned in a way identical to that of observed CDDF. Let $f_{{\rm sim},i,k}(\Gamma_{\rm 12})$ be value of the CDDF  in $k^{th}$ bin of $i^{\rm th}$ mock sample for a given \GTW. The mean and variance of CDDF in $k^{th}$ bin  is calculated  as follows.
\begin{equation}\label{eq:mean-sigma-cdd}
\begin{aligned}
\overline{f}_{{\rm sim},k}(\Gamma_{\rm 12}) &= \frac{1}{N} \sum \limits_{i=1}^{N} f_{{\rm sim},i,k}(\Gamma_{\rm 12}) \\ 
\sigma^2_{{\rm sim},k} &= \frac{1}{N-1} \sum \limits_{i=1}^{N}  [f_{{\rm sim},i,k}(\Gamma_{\rm 12}) - \overline{f}_{{\rm sim},k}(\Gamma_{\rm 12})]^2
\end{aligned}
\end{equation}
where $N=100$ is number of mock samples. \\
The reduced $\chi^2$ between observed CDDF and model CDDF is given as,
\begin{equation}
\begin{aligned}
\chi^2_{\rm dof}(\Gamma_{\rm 12}) = \frac{1}{N_{bin} -1} \sum \limits_{k=1}^{N_{bin}} \frac{[\overline{f}_{{\rm sim},k}(\Gamma_{\rm 12}) - f_{{\rm obs},k}]^2}{\sigma^2_{{\rm sim},k}}
\end{aligned}
\end{equation}
where $N_{bin} = 13$ is number of bins in CDDF and $f_{{\rm obs},k}$ is observed CDDF in $k^{\rm th}$ bin.

Right-hand panels of Fig. \ref{fig:gamma-12-constraint} show the constraints on \GTW from CDDF for a model without turbulence in the $4$ different redshift bins. As expected the $\chi^2$ curves are smooth parabolas with the minimum reduced $(\chi^2_{\rm dof,min})$ $\sim 1.5$ as given in Table \ref{tab:error-budget}. The statistical uncertainty in \GTW is shown by black dashed vertical line. The statistical uncertainty in \GTW is calculated by demanding $\chi^2 = \chi^2_{\rm min} + \Delta \chi^2$, where $\chi^2_{\rm min}$ is minimum value of $\chi^2$ and $\Delta \chi^2 = 1.0$ for $1$ degree of freedom \citep[corresponds to 1 free parameter in the problem i.e., \GTW; ][]{press1992}. The shaded regions shown in the right-hand panels correspond to $1\sigma$ constraint on \GTW from \citetalias{gaikwad2016}.
It is interesting to note that the \GHI constraints obtained using CDDF in this work (see Table \ref{tab:error-budget}) are consistent within $\sim 1\sigma$ (the best fit values differ by $< 5.5$ percent) with those obtained using flux statistics in \citetalias{gaikwad2016} (see Table 8 in \citetalias{gaikwad2016}). However, error range is smaller in the present study. It is because, contrary to \citetalias{gaikwad2016}, here we do not consider the errors arising from uncertainties in cosmological parameters, continuum fitting and cosmic variance. Therefore, we caution the reader that errorbars on \GTW from \citetalias{gaikwad2016} are more realistic.

Left-hand panels in Fig. \ref{fig:gamma-12-constraint} shows the best fit (i.e., models with \GTW corresponding to minimum $\chi^2_{\rm dof}$) model CDDF (blue square with shaded region)  and observed CDDF (red circle with errorbar) in the $4$ redshift bins. The gray shaded region represents $1\sigma$ range from the simulated mock sample ($\sigma_{\rm sim}$ given in Eq.\ref{eq:mean-sigma-cdd}). The errorbars shown on observed CDDF are assumed to be poisson distributed and are not used while calculating $\chi^2$.

The redshift bin IV is likely to be affected by Ly-$\beta$ contamination from H~{\sc i} interlopers (refer to \citetalias{danforth2016} and \citetalias{gaikwad2016}). Note that it is difficult to remove the Ly-$\beta$ contamination in observations due to limited wavelength coverage of the spectrograph. We accounted for this effect by contaminating the simulated Ly-$\alpha$ forest in the redshift $0.35 \leq z \leq 0.45$ with Ly-$\beta$ forest from simulation box at $z=0.6$. (for more detail we refer reader to \S5.1 in \citetalias{gaikwad2016}). The lowest panels in Fig. \ref{fig:gamma-12-constraint} shows the \GTW constraint at $z=0.4$ where model CDDF is calculated from Ly-$\alpha$ forest contaminated with Ly-$\beta$ forest. The  $\chi^2_{\rm dof,min} = 1.35$ and \GTW constraint are consistent  (within $1\sigma$) with best fit \GTW from \citetalias{gaikwad2016} ($0.210 \pm 0.052$) in the same redshift bin\footnote{If we do not account for Ly-$\beta$ contamination, the  \GTW constraint is underestimated, a result similar to \citetalias{gaikwad2016}. For Ly-$\alpha$ forest only simulation at $z=0.4$, $\Gamma_{\rm 12} = 0.176 \pm 0.013$}.

We have done similar analysis and constrained \GTW for a model with Gaussian distributed $b_{\rm turb}$ (see \S\ref{sec:b-parameter-distribution}) as shown in Fig. \ref{fig:gamma-12-constraint-turb} (see Table \ref{tab:error-budget} also). It is interesting to note that the \GTW constraints obtained from the model with Gaussian distributed $b_{\rm turb}$ are in good agreement with those from model without turbulence and \citetalias{gaikwad2016} (see left-hand panel in Fig. \ref{fig:gamma-12-evolution}). This suggests that the addition of Gaussian distributed $b_{\rm turb}$ has a mild effect on CDDF (and hence \GTW constraints). The $\chi^2_{\rm dof,min}$ is also improved and close to 1 when Gaussian distributed $b_{\rm turb}$ is added to the model.

We summarize the best fit \GTW (for the model with and without turbulence) along with the statistical errors for $4$ redshift bins in Table \ref{tab:error-budget}. It is clear from left-hand panel of Fig. \ref{fig:gamma-12-evolution} that there is evolution in \GTW with redshift and that follows $\Gamma_{\rm HI} (z) = (3.9 \pm 0.1) \times 10^{-14} \; (1+z)^{4.98 \pm 0.11} \; {\rm s^{-1}}$ up to $z = 0.5$. 

\subsection{Implications of derived \GHI:}
\label{subsec:gamma-12-implications}
In this section we compare our derived \GHI with our previously measured \GHI from \citetalias{gaikwad2016} and \GHI from the literature at low-$z$. Note previous studies (except \citetalias{gaikwad2016}) do not get \GHI through $\chi^2$ minimization and no errorbars are associated to the \GHI measurement. The \GTW evolution from this work using CDDF for models with and without turbulence is in good agreement with \GTW evolution (using flux PDF and flux PS) from \citetalias{gaikwad2016}. The uncertainty in \GTW constraint from \citetalias{gaikwad2016} is larger and more reliable as they also account for other systematic and statistical uncertainties.

Right-hand panel in Fig. \ref{fig:gamma-12-evolution} shows that our derived \GTW ($0.066 \pm 0.06$) at $z=0.1$ is lower than \GTW from \citetalias{kollmeier2014} ($\sim 0.178$) by factor $\sim 2.7$ but is consistent with \citetalias{shull2015} ($\sim 0.070$) within $1\sigma$. \citetalias{kollmeier2014} compared the CDDF in the redshift range $0 \leq z \leq 0.75$ with models generated from simulation box at $z_{\rm sim}=0.1$. This could be the possible reason for the discrepancy of \GTW constraints obtained in this work and \citetalias{kollmeier2014}. 
We constrained \GTW from CDDF in $0.075 \leq z \leq 0.45$ using models generated from simulation box at $z_{\rm sim}=0.1$ ignoring evolution in \GTW and large scale structures.
The \GTW constraint ($0.092 \pm 0.009$) for this model is higher by factor $\sim 1.4$ as compare to \GTW constraint from the model at $z=0.1$ where \GTW evolution is accounted for (see Table \ref{tab:error-budget}). However, the \GTW constraint from this model is in good ($\sim 1\sigma$) agreement with \GTW constraint at $z=0.2$ ($0.104 \pm 0.008$, see Table \ref{tab:error-budget}) when \GTW evolution is accounted for. This is still smaller than what was found by \citetalias{kollmeier2014}. Note that the median redshift of the observed sample is $z \sim 0.2$.
Thus even if we do not account for \GTW evolution, the \GTW constraints ($0.092 \pm 0.009$) are consistent with \citetalias{shull2015} ($\sim 0.103$ at $z=0.2$) and \citetalias{gaikwad2016} ($0.100 \pm 0.021$ at $z=0.2$). 

Recently, \citet{gurvich2016} found that the \GTW required to match observed CDDF (from \citetalias{danforth2016}) with simulated CDDF is lower than \citetalias{kollmeier2014} by a factor $\sim 3$. They attributed this to the effect of AGN feedback and to the \citet{faucher2009} UVB model  included in their simulation. The AGN feedback could be important but is not incorporated in our simulation. However our \GTW constraints are in good agreement with those from \citet{gurvich2016}. Note that our \GTW constraint at $z=0$ (\GTW $ = 0.039 \pm 0.001$, from scaling relation) is consistent within $1\sigma$ with that from \citet{faucher2009} UVB model (\GTW $ = 0.0384$).
\citet{viel2016} have provided scaled \GTW for different hydrodynamical simulations with and without additional heating that will reproduce the observed CDDF. The direct comparison between our \GTW measurements and that of \citet{viel2016} is difficult as goodness of the CDDF fits and error in \GTW are not given in their work. However if we assign 10 percent uncertainty to their measurements, they are very much in agreement with our measurements.
\citet{cristiani2016} found a similar result where their estimated \GTW evolution at $z < 0.5$ is in good agreement with \GTW evolution from \citetalias{shull2015} and \citetalias{gaikwad2016}. \citet{cristiani2016} combined the information from measured QSO contribution to cosmic UVB in the range $3.6 < z < 4.0$ with QSO luminosity function and estimated the production of ionizing photons from QSOs at various epoch ($0 \leq z \leq 5$). 

We compare our measured \GHI with the UVB model. We find that the UVB generated using updated QSO emissivity \citep{khaire2015b} is in very good agreement with our \GHI measurements. The gray shaded region in right-hand side of Fig. \ref{fig:gamma-12-evolution} shows the \GHI obtained using this UVB model. This shaded region accounts for two different H~{\sc i} distributions \citep[from][]{haardt2012,inoue2014} and spectral energy distributions of QSOs \citep{stevans2014,lusso2014} used in the UVB models. \GHI measurements can be used to place constraints on escape fraction ($f_{\rm esc}$) of the H~{\sc i} ionizing photons from galaxies \citep[see][]{inoue2006,khaire2016}. Using \GHI measured here and uncertainties on it from \citetalias{gaikwad2016}, we find that the $f_{\rm esc}$ is negligible \citep[with star formation history from][the $3\sigma$ limits are 0.8 percent]{khaire2015a}, consistent with measurements of average $f_{\rm esc}$ from low-$z$ galaxies \citep{cowie2009,bridge2010,siana2010,rutkowski2015}. Therefore, low-$z$ UVB is predominantly contributed by QSOs.

\InputFigCombine{Gamma_12_evolution_K14_S15.pdf}{170}{%
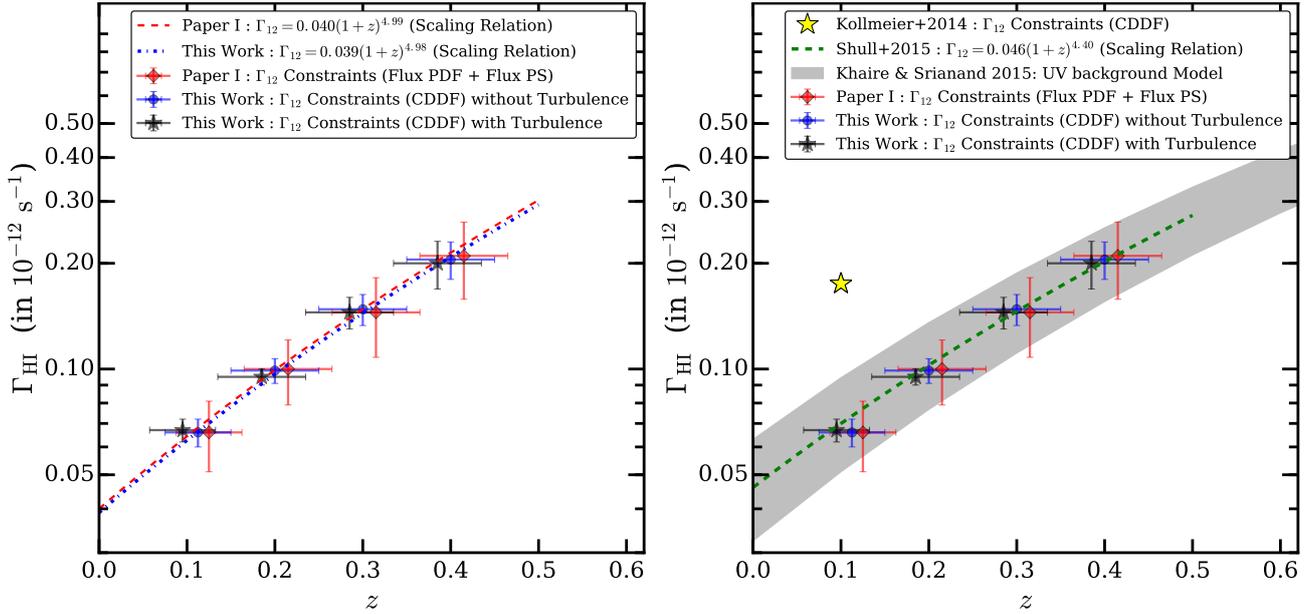
In the left-hand panel black stars and blue circles with errorbar show the evolution of \GTW from this work (CDDF) for a model with turbulence (see \S\ref{sec:b-parameter-distribution}) and without turbulence respectively. Red diamonds with errorbars  show the \GTW evolution from \citetalias{gaikwad2016} (using flux PDF and flux PS). For visual purpose the points are shifted along $x$ axis. The blue dotted and red dashed line shows scaling relation for \GTW evolution from this work (for model without turbulence) and \citetalias{gaikwad2016}. Within $1\sigma$ uncertainty the \GTW evolution from this work (for both models with and without turbulence) is consistent with that from \citetalias{gaikwad2016}. The errorbars on \GTW evolution given in \citetalias{gaikwad2016} are more realistic as they account for cosmic variance, cosmological parameter uncertainty, continuum fitting uncertainty. In the right-hand panel black stars, blue circles and red diamonds with errorbars are same as given in left-hand panel. Our derived \GTW at $z\sim0.1$ is lower by factor $\sim 2.7$ than that of \citetalias{kollmeier2014} (shown by yellow star) but is in $1 \sigma$ agreement with \citetalias{shull2015} (green dashed line). The derived \GHI evolution is consistent with \citet{khaire2015b,khaire2015a} UVB model (shown by gray shaded region) where the UVB is contributed only by QSOs. }{\label{fig:gamma-12-evolution}}


\begin{table*}
\caption{\GTW measurements at different redshifts}
\begin{threeparttable}
\centering
\begin{tabular}{lccccc}
\hline \hline
Redshift bin $\Rightarrow$ & I & II & III & IV \\ 
Type of simulated spectra $\Rightarrow$ & Ly-$\alpha$ forest & Ly-$\alpha$ forest & Ly-$\alpha$ forest & Ly-$\alpha$ + Ly-$\beta$ forest\tnote{1}  \\ \hline \hline
Best Fit \GTW (without turbulence) & 0.066 & 0.104 & 0.137 & 0.199 \\ 
Statistical uncertainty\tnote{2} & $\pm$0.006 & $\pm$0.008 & $\pm$0.015 & $\pm$0.025 \\  
Reduced $\chi^2_{\rm min}$ & 1.49 & 1.39 & 1.24 & 1.35 \\ \hline 
Best Fit \GTW (for Gaussian $b_{\rm turb}$)& 0.067 & 0.095 & 0.145 & 0.200 \\ 
Statistical uncertainty\tnote{2} & $\pm$0.005 & $\pm$0.005 & $\pm$0.015 & $\pm$0.022 \\  
Reduced $\chi^2_{\rm min}$ & 1.26 & 1.09 & 1.04 & 1.00 \\ \hline \hline
\end{tabular} 
\begin{tablenotes}
\item[1] Following \citetalias{gaikwad2016}, simulated Ly-$\alpha$ forest at $z=0.35$ to $0.45$ is contaminated by Ly-$\beta$ forest in the same wavelength range. The Ly-$\beta$ forest is generated from simulation box at $z=0.6$ . 
\item[2] The uncertainty in \GTW due to uncertainty in thermal history parameters is well within statistical uncertainty.
\end{tablenotes}
\end{threeparttable}
\label{tab:error-budget}
\end{table*}

\section{Summary}
\label{sec:summary}
In this work we constrain the H~{\sc i} photoionization rate  \GHI and its redshift evolution at $z <0.45$ from a sample of 82 QSO obtained from Cosmic Origins Spectrograph on board Hubble Space Telescope using H~{\sc i} column density distribution function (CDDF). To explore full \GHI range we have developed a code ``\fullcodename (\codename)'' to automatically fit the Ly-$\alpha$ absorption lines with Voigt profile. This code is parallel and is written in \textit{python}. The main results of this work are as follows
\begin{itemize}
\item We fitted all the observed Ly-$\alpha$ forest spectra using \codename and compiled a Ly-$\alpha$ line catalog called ``\codename line catalog''. The fitted parameters such as column density (${\rm N_{HI}}$), line width parameter ($b$) and line width distribution from \codename line catalog are found to be consistent with those from \citet{danforth2016}.
The median $b$ parameter from {\sc viper} ($32.9 \pm 20.8$ \kmps) is consistent with that from \citet[][$33.9 \pm 18.3$ \kmps]{danforth2016}. Whereas, the median \logNHI from {\sc viper} ($13.39 \pm 0.61$) is in good agreement with that from \citet[][$13.38 \pm 0.63$]{danforth2016}
\item We calculate the appropriate redshift path length  $\Delta z ({\rm N_{HI}})$ and the sensitivity curve from {\sc hst-cos} data. We calculate the CDDF after accounting for the incompleteness of the sample. Our calculated CDDF in the redshift range ($0.075 \leq z \leq 0.45$) is consistent (KS test $p$-value is 0.83) with that of \citet{danforth2016} CDDF in the redshift range ($0 \leq z \leq 0.75$). 
\item We found that the $b$ parameters of Voigt profile components from simulations are typically underestimated as compared to observations. This difference can be rectified by including the Gaussian distributed line width parameter $b_{\rm turb}$ ($\mu=20$ \kmps and $\sigma=10$ \kmps) at each pixel in the simulation. The resulting line width distribution from simulations matches roughly with observed line width distribution, scatter and lower envelope of the $b$ vs \logNHI distribution. However, the CDDF has little effect of additional $b_{\rm turb}$  ($< 7$ percent) and the \GHI constraints are mildly affected  ($< 9$ percent). However, if we consider additional heating effect for the excess broadening then the \GHI obtained will be slightly reduced (roughly scale as $T^{-0.7}$).
\item We obtained CDDF at four different $z$ bins and matched with simulated CDDF at each mean $z$. This allowed us to measure the \GHI in four redshift bins (of $\Delta z = 0.1$) centered around $\overline{z} = 0.1125,0.2,0.3,0.4$. We estimated the  associated statistical error using $\chi^2$ statistics. When additional turbulent broadening are not included measured \GTW values at the redshift bins $\overline{z} = 0.1125,0.2,0.3,0.4$ are \GTW $=0.066 \pm 0.006, 0.104 \pm 0.008, 0.137 \pm 0.015, 0.199 \pm 0.025$ respectively. The corresponding values after inclusion of $b_{\rm turb}$ are  \GTW $=0.067 \pm 0.005, 0.095 \pm 0.005, 0.145 \pm 0.015, 0.200 \pm 0.022$. Thus the uncertainties in the velocity broadening seem to have little effect on the derived \GHI. Our measured \GTW values are in good agreement with \citet{gaikwad2016} \GTW measurement that are obtained with two different statistics namely flux PDF and flux PS. However, the errorbars on \GHI measurements of \citet{gaikwad2016} are more reliable as they account for cosmic variance, continuum fitting uncertainty and cosmological parameter uncertainty in their calculation.
\item Our measured \GHI is increasing with $z$ and follows the relation $\Gamma_{\rm HI}(z)=(3.9 \pm 0.1) \times 10^{-14} \; (1+z)^{4.98 \pm 0.11} \;{\rm s^{-1}}$ which is in agreement with \citet{shull2015}. The measured \GHI evolution is consistent with \citet{khaire2015b,khaire2015a} UV background model where it is contributed only by QSOs (i.e., the galaxy contribution is negligible). 
Inclusion of additional thermal broadening will reduce the \GHI value obtained. This will further reduce the galaxy contribution to the \GHI at low-$z$.
\end{itemize}

\vspace{-9mm}
\section*{Acknowledgment}
All the computations were performed using the PERSEUS cluster at IUCAA and the HPC cluster at NCRA. We would like to thank Volker Springel, Matteo Viel and Simeon Bird for useful discussion. We also like to thank the referee John Webb and Matthew Bainbridge for improving this work and manuscript.



\bibliographystyle{mnras}
\bibliography{VIPER} 

\section*{Appendix}
We formed a line catalog by fitting the observed spectra using \codename. Table \ref{tab:viper-line-catalog} shows few fitted parameters from the \codename line catalog for spectra towards QSO 1ES1553+113. The first, second, third, fourth, fifth and sixth column shows fitted wavelength ($\lambda$ in ${\rm \AA}$), error in wavelength ($d\lambda$ in ${\rm \AA}$), log of column density (\logNHI in \pcmsq), error in log of column density (\dlogNHI in \pcmsq), $b$ parameter (in \kmps) and error in $b$ parameter ($db$ in \kmps) respectively. The full \codename line catalog is available online in ASCII format with this paper. 
\begin{table}
\caption{Few fitted parameters from the \codename line catalog are given for spectra towards QSO 1ES1553+113. The full line catalog is available online in ASCII format.}
\centering
\begin{tabular}{cccccc}
\hline \hline
$\lambda$ & d$\lambda$ & \logNHI & \dlogNHI & $b$  & $db$  \\ 
(${\rm \AA}$) & (${\rm \AA}$) & (\pcmsq) & (\pcmsq) & (\kmps) & (\kmps)   \\  \hline \hline
1330.827 & 0.003 & 13.68 & 0.01 & 32.06 & 1.08 \\ 
1339.936 & 0.002 & 14.23 & 0.01 & 38.09 & 0.80 \\ 
1361.466 & 0.019 & 12.84 & 0.06 & 34.10 & 7.00 \\ 
1361.922 & 0.013 & 12.94 & 0.05 & 28.91 & 4.99 \\ 
1365.408 & 0.018 & 13.04 & 0.04 & 48.75 & 5.57 \\ \hline
\end{tabular}
\label{tab:viper-line-catalog}
\end{table}




\bsp	
\label{lastpage}
\end{document}